\documentclass{ws-ijmpa}
\usepackage[super,compress]{cite}
\usepackage{graphicx}
\usepackage[T1]{fontenc} 
\usepackage{mathrsfs}

\usepackage{color}

\begin{document}
\markboth{J.H. Gao, Z.T. Liang, Q. Wang}{Quantum kinetic theory in Wigner function formalism}

%
\catchline{}{}{}{}{}
%

\title{Quantum kinetic theory for spin-1/2 fermions in Wigner function formalism
}

\author{Jian-Hua Gao
}

\address{Shandong Provincial Key Laboratory of Optical Astronomy and Solar-Terrestrial
Environment, Institute of Space Sciences, Shandong University\\
Weihai, Shandong 264209, China
\\
gaojh@sdu.edu.cn}

\author{Zuo-Tang Liang}
\address{Institute of Frontier and Interdisciplinary Science,
Key Laboratory of Particle Physics and Particle Irradiation (MOE), Shandong University\\
Qingdao, Shandong 266237, China\\
liang@sdu.edu.cn}

\author{Qun Wang}

\address{Department of Modern Physics, University of Science and Technology of China\\
Hefei, Anhui 230026, China\\
qunwang@ustc.edu.cn}

\maketitle


\begin{abstract}
{
We give a brief overview of the kinetic theory for spin-$1/2$ fermions { in Wigner function formulism}. }
The chiral and spin kinetic equations can be derived from equations for Wigner functions.
A general Wigner function has 16 components which satisfy 32 coupled equations.
For massless fermions, the number of independent equations can be significantly
reduced due to the decoupling of left-handed and right-handed particles.
It can be proved that out of many components of Wigner functions and their coupled equations,
only one kinetic equation for the distribution function is independent.
This is called the disentanglement theorem for Wigner functions of chiral fermions.
For massive fermions, it turns out that one particle distribution function
and three spin distribution functions are independent and satisfy four kinetic equations.
Various chiral and spin effects such as chiral magnetic and votical effects, the chiral seperation effect,
spin polarization effects can be consistently described in the formalism.

\end{abstract}

\ccode{PACS numbers:}


\section{Introduction}
Relativistic heavy-ion collisions provide a unique chance to create a new state of strong interaction matter called the quark-gluon plasma (QGP) which is a deconfined phase of nuclear matter under extreme conditions of high temperatures and densities. The complexity in relativistic heavy-ion collisions lies in the fact that the dynamical processes happen at very small space-time scale of the order 10 fm. Global properties of relativistic heavy-ion collisions can be well described by thermodynamic and hydrodynamic models which assume fast local equilibrium after collisions. However non-equilibrium and quantum effects may play an important role in such violent and microscopic processes at small time-scale. For example, coherent gluon fields called the color glass condensate may provide an initial condition for the formation of QGP far from equilibrium\cite{Gribov:1984tu,Mueller:1985wy,McLerran:1993ka,McLerran:1994vd,Iancu:2003xm} .
To investigate how these effects influence experimental observables is an  important task which can be achieved with the help of the quantum transport theory. 	

Since 1980s, there have been tremendous efforts in constructing a quantum transport theory for the QGP in order to describe the non-equilibrium and quantum phenomena \cite{Heinz:1983nx,Winter:1984ztp,Heinz:1985vf,Heinz:1984my,Heinz:1984yq,Gyulassy:1986jq,Heinz:1984uc,Winter:1986da,Heinz:1985qe,Elze:1986qd
,Elze:1986hq,Heinz:1986kz,Vasak:1987um,Elze:1987ii,Calzetta:1987bw,Fonarev:1990ph,Weigert:1990qv,Mrowczynski:1992hq,Fonarev:1993ht,
Zhuang:1995pd,Zhuang:1995jb,Abada:1996hq,Antonsen:1997dc,Ochs:1998qj,Zhuang:1998bqx,Holl:2001fs,Morozov:2002hv,Prozorkevich:2003td,Yamamoto:2003yy}.
These works can be grouped into those about
Abelian gauge fields~\cite{Vasak:1987um,Fonarev:1990ph,Zhuang:1995pd,Zhuang:1995jb,Abada:1996hq,Zhuang:1998bqx,Holl:2001fs,Morozov:2002hv}
and non-Abelian gauge fields~\cite{Heinz:1983nx,Winter:1984ztp,Heinz:1985vf,Heinz:1984my,Heinz:1984yq,Gyulassy:1986jq,Heinz:1984uc,Heinz:1985qe,Elze:1986qd,Elze:1986hq,Heinz:1986kz,Weigert:1990qv,Fonarev:1993ht,Ochs:1998qj}.
Most of these works are based on covariant Wigner functions while a part of them are formulated in equal-time Wigner functions~\cite{Zhuang:1995pd,Zhuang:1995jb,Abada:1996hq,Ochs:1998qj,Zhuang:1998bqx,Holl:2001fs,Morozov:2002hv,Prozorkevich:2003td}.
For applications in astrophysics a generalization of Wigner functions in the curved space-time was made~\cite{Winter:1986da,Fonarev:1990ph,Fonarev:1993ht,Antonsen:1997dc,Yamamoto:2003yy}.
The reviews on these earlier devolvements can be found in Refs.~\citen{Elze:1989un},~\citen{Blaizot:2001nr}.
These works was done in the era before RHIC was run at BNL.
However, the Wigner function formalism has not been put into simulations in rigorous way to describe data at RHIC or LHC.
One reason is that the formalism is rather complicated in which multiple components are entangled with each other and the transport equation are highly constrained.
This made it very difficult to be solved in numerical simulation.
Another reason is that most results on collective flows at RHIC or LHC energies can be well described by relativistic hydrodynamics
with proper initial conditions~\cite{Heinz:2013th,Gale:2013da,Teaney:2009qa,Romatschke:2007mq,Song:2007fn,Dusling:2007gi,Molnar:2008xj,Schenke:2010rr}.

In recent years, the theoretical prediction~\cite{Liang:2004ph,Liang:2004xn,Gao:2007bc}
and the experimental observation by STAR Collaboration at RHIC~\cite{STAR:2017ckg,Adam:2018ivw} of global polarization effect (GPE)
have drawn great attention in the field, both experimentally and theoretically (see e.g. recent short reviews~\cite{Gao:2020lxh,Gao:2020vbh,Becattini:2020ngo}).
A series of other quantum effects in related aspects in heavy ion collisions have been extensively studied, such as the chiral magnetic effect (CME)~\cite{Vilenkin:1980fu,Kharzeev:2007jp,Fukushima:2008xe},
the chiral vortical effect (CVE)~\cite{Vilenkin:1978hb,Kharzeev:2007tn,Erdmenger:2008rm,Banerjee:2008th}, the chiral separation effect (CSE)~\cite{Son:2004tq,Metlitski:2005pr},
and the local polarization effect (LPE)~\cite{Gao:2012ix}.
This opens a new window to probe the nature of the QGP in a different perspective.
Since all these effects are associated with the particle's spin degrees of freedom, the conventional relativistic hydrodynamics or Vlasov-Boltzmann equation have to be generalized to describe these quantum effects in a consistent way.
The quantum transport theory is the appropriate tool for this goal.
%
%
%


The quantum kinetic theory based on Wigner functions was proposed to derive the CME, CVE, CSE and LPE in the Ref.~\citen{Gao:2012ix},
which is a success in describing these effects in a natural and consistent way.
Since then, there have been a large number of works along this line on various chiral or spin effects in relativistic heavy ion collisions, condensed matter physics, and astrophysics.
In this article, we will give an overview for recent developments of the Wigner function formalism \cite{Chen:2012ca,Chen:2013dca,Gao:2015zka,Guo:2017dzf,Hidaka:2016yjf,Fang:2016vpj,Huang:2018wdl,
Gao:2018wmr,Hidaka:2017auj,Hidaka:2018ekt,Gao:2017gfq,Gao:2018jsi,Liu:2018xip,Gao:2020ksg,Gao:2019znl,
Sheng:2018jwf,Weickgenannt:2019dks,Hattori:2019ahi,Wang:2019moi,
Dayi:2019hod,Lin:2019fqo,Liu:2020flb,Tabatabaee:2020efb,Gao:2019zhk,Sheng:2020oqs,Dayi:2020uwx,Guo:2020zpa,Huang:2020wrr,Shi:2020htn,Florkowski:2017dyn,Florkowski:2018ahw,
Bhadury:2020puc,Yang:2020mtz,Hou:2020mqp}.
Especially, we will focus on (a) Disentanglement of covariant Wigner functions into  a single distribution function which satisfies a single kinetic equation for massless fermions
{or
 one distribution function and three spin distribution functions which satisfy four kinetic equations for massive fermions} in background electromagnetic fields; (b)Derivation of various chiral and spin effects such as chiral anomaly, CME, CVE, etc..  {There are a number of interesting topics that we do not cover in this review article, e.g.,
the smooth transition from the spin kinetic theory for massive fermions to chiral kinetic theory for massless fermions \cite{Weickgenannt:2019dks,Hattori:2019ahi,Wang:2019moi,Sheng:2020oqs,Guo:2020zpa}, the quantum kinetic theory in curved space \cite{Winter:1986da,Calzetta:1987bw,Fonarev:1990ph,Fonarev:1993ht,Antonsen:1997dc,Yamamoto:2003yy,Liu:2018xip,Liu:2020flb},
the chiral kinetic equation in relaxation time approximation \cite{Hidaka:2017auj,Hidaka:2018ekt}, the chiral kinetic equation in strong magnetic field \cite{Lin:2019fqo,Gao:2020ksg},
the chiral kinetic equation with vorticity or  in rotating frame \cite{Chen:2012ca,Dayi:2019hod,Dayi:2020uwx,Hou:2020mqp,Huang:2020kik}, and derivation of the spin hydrodynamics from Wigner equations \cite{Florkowski:2017dyn,Florkowski:2018ahw,Bhadury:2020puc,Shi:2020htn}.
Also the numerical simulation of the chiral kinetic equation \cite{Sun:2016nig,Sun:2016mvh,Zhou:2018rkh,Liu:2019krs} is not discussed.
We refer readers who are interesting in these topics to the references listed above.
In addition to the Wigner function approach, } these chiral and spin effects can also be studied in other approaches, such as the AdS/CFT duality
\cite{Newman:2005hd,Yee:2009vw,Rebhan:2009vc,Gorsky:2010xu,Gynther:2010ed,Eling:2010hu,Hoyos:2011us,Amado:2011zx,
Nair:2011mk,Kalaydzhyan:2011vx,Lin:2013sga}, relativistic anomalous and spin hydrodynamics
\cite{Son:2009tf,Pu:2010as,Lublinsky:2009wr,Sadofyev:2010pr,Neiman:2010zi,Lin:2011mr,Bhattacharya:2011tra,
Kharzeev:2011ds,Hattori:2019lfp,Montenegro:2017rbu,Florkowski:2017ruc,Florkowski:2018myy,Becattini:2018duy},
quantum field theories
\cite{Kharzeev:2007jp,Fukushima:2008xe,Kharzeev:2009pj,Fukushima:2009ft,Asakawa:2010bu,Fukushima:2010vw,
Fukushima:2010zza,Landsteiner:2011cp,Hou:2011ze,Hou:2012xg,Becattini:2015nva,Buzzegoli:2017cqy,Buzzegoli:2018wpy,Lin:2018aon,Feng:2018tpb,Dong:2020zci}.
The chiral kinetic equation can also be derived from other methods such as semiclassical approaches~\cite{Duval:2005vn,Wong:2011nt,Son:2012wh,Stephanov:2012ki,Dwivedi:2013dea,Chen:2014cla,Manuel:2014dza,Huang:2018aly},
effective field theories \cite{Son:2012zy,Carignano:2018gqt,Lin:2019ytz,Carignano:2019zsh},
and the world-line formalism~\cite{Mueller:2017lzw,Mueller:2017arw,Mueller:2019gjj}.
The reviews on these chiral and spin effects can be found in Refs.~\citen{Kharzeev:2013ffa,Huang:2015oca,Kharzeev:2015znc,Liu:2020ymh,Gao:2020vbh} .

The article is organized as follows. In Section \ref{sec-Wigner}, we review the Lorentz covariant and gauge invariant quantum transport theory for spin-1/2 fermions in background Abelian gauge fields. The spinor decomposition and semiclassical expansion are presented. This is the starting point for the following sections. In Section \ref{sec-disentangle}, we give a systematic method to disentangle multicomponent Wigner funcitons and their equations into independent distribution functions and their kinetic equations so that the number of degrees of freedom is greatly reduced. We will discuss the derivation of Wigner functions for massless and massive fermions separately. In Section \ref{sec-chiral}, we present a consistent way of describing various chiral and spin effects in covariant Wigner functions. The summary and outlook are made in the final section.

We adopt following notation and conventions: $a\cdot b=a^{\mu}b_{\mu}$, $g_{\mu\nu}=\mathrm{diag}(+,-,-,-)$,
$\epsilon^{0123}=-\epsilon_{0123}=1$, $\gamma _5=\gamma ^5=i\gamma^0\gamma^1\gamma^2\gamma^3$,
$\sigma^{\mu\nu}=(i/2)[\gamma^\mu,\gamma^\nu]$, and summation over repeated
indices is implied if not stated explicitly.


\section{Wigner function formalism}
\label{sec-Wigner}
The classical transport theory describes the time evolution of the distribution $f(t,{\bf x},{\bf p})$,
a measure of the particle number density in phase space, which satisfies the Vlasov-Boltzmann equation,
\begin{eqnarray}
\label{Boltzmann-eq}
\partial_t f(t,{\bf x},{\bf p})  +  {\bf v} \cdot {\pmb\nabla}_x f(t,{\bf x},{\bf p})
+  \left( {\bf E} +{\bf v}\times {\bf B} \right)\cdot {\pmb \nabla}_{p} f(t,{\bf x},{\bf p}) = {\cal C} [f]\,,
\end{eqnarray}
 where ${\bf v}={\bf p}/E_p$ denotes the particle's velocity with mass $m$ and energy $E_p=\sqrt{{\bf p}^2+m^2}$, ${\bf E}$ and ${\bf B}$ are the electric field and magnetic field respectively, and ${\cal C} [f]$ denotes collision terms.
For notational simplicity, we have already absorbed the electric charge `$e$' into ${\bf E}$ and ${\bf B}$.

In quantum theory, we use the Wigner function $W(x,p)$, a quantum anologue of $f(t,{\bf x},{\bf p})$, first introduced by E. Wigner in Ref.~\citen{Wigner:1932eb}. In quantum electrodynamics, the Lorentz covariant Wigner function is defined as an ensemble average of the Wigner operator \cite{Vasak:1987um}
\begin{eqnarray}
\label{wigner}
 W_{\alpha\beta}(x,p)&=&\int\frac{d^4 y}{(2\pi)^4}
e^{-ip\cdot y}\left\langle \bar\psi_\beta(x+\frac{y}{2})U(x+\frac{y}{2},x-\frac{y}{2})
\psi_\alpha(x-\frac{y}{2})\right\rangle\,,
\end{eqnarray}
where $\psi_{\alpha}$ and $\bar\psi_{\beta}$ are Dirac fields with $\alpha,\beta$ running from 1 to 4 in spinor space and the phase factor $U$ is the so-called Wilson line or gauge link
\begin{eqnarray}
\label{link}
U(x+\frac{y}{2},x-\frac{y}{2})&\equiv&
{\cal P}e^{-iy^\mu\int_0^1 ds A_\mu\left(x-\frac{y}{2}+sy\right)}\,,
\end{eqnarray}
which ensures the gauge invariance of the Wigner function.
The operator ${\cal P}$ denotes a certain path ordering with respect to the
parameter $s$. Under the requirement that the Wigner function measures the particle density with the physical kinetic momentum $p$, the path must be a straight line. Again we have absorbed the electric charge into the gauge
potential $A_\mu$.
For simplicity,  we will restrain ourselves to the case of classical background fields
that the gauge potential $A_\mu$ and field strength tensor $F_{\mu\nu}$ are classical functions of time and space.
In this case the path ordering operator ${\cal P}$ is just a unit operator.
In the background field approximation, there is no issue of the so-called BBGKY-hierarchy \cite{DeGroot:1980dk}.

{We note in particular that
we define the ensemble average $\langle\cdots\rangle$ in Eq.~(\ref{wigner}) without taking the normal ordering of Dirac fields
because the Dirac equation is satisfied by the Wigner function, but the normal ordering will destroy such a property.
This is crucial~\cite{Gao:2019zhk} to reproduce the correct and universal coefficient of chiral anomaly and we will come to this point in Sec.~\ref{chiral-anomaly}.}

{In this section, we give a brief review  on the main results presented in Ref.~\citen{Vasak:1987um},
which  is the starting point for the quantum transport theory we are going to address in this review article.}
The covariant Wigner function is a $4\times 4$ matrix and
can be expanded in terms of the 16 independent generators of
the Clifford algebra
\begin{eqnarray}
\label{decomposition}
W=\frac{1}{4}\left[\mathscr{F}+i\gamma^5 \mathscr{P}+\gamma^\mu \mathscr{V}_\mu +\gamma^5 \gamma^\mu\mathscr{A}_\mu
+\frac{1}{2}\sigma^{\mu\nu} \mathscr{S}_{\mu\nu}\right]\,,
\end{eqnarray}
where all coefficients are real functions due to the property $W^\dagger=\gamma^0 W \gamma^0$ and can be determined by taking traces with proper generators
\begin{eqnarray}
\mathscr{F}(x,p)&=&\textrm{tr}\left[ W(x,p)\right],\\
\mathscr{P}(x,p)&=& -i\textrm{tr}\left[\gamma^5 W(x,p)\right],\\
\mathscr{V}_\mu(x,p)&=& \textrm{tr}\left[\gamma_\mu W(x,p)\right],\\
\mathscr{A}_\mu(x,p)&=& \textrm{tr}\left[\gamma_\mu \gamma^5 W(x,p)\right],\\
\mathscr{S}_{\mu\nu}(x,p)&=& \textrm{tr}\left[\sigma_{\mu\nu} W(x,p)\right].
\end{eqnarray}
In background fields, the equations of motion for Wigner functions can be derived from the Dirac equation,
\begin{eqnarray}
\label{eq-c}
\left(\gamma\cdot K -m\right) W(x,p)=0\,,
\end{eqnarray}
where the operators $K^\mu$, $G^\mu$ and $\Pi^\mu$ are defined by
\begin{eqnarray}
\label{Kmu}
K^\mu &\equiv& \Pi^\mu +\frac{1}{2}i\hbar G^\mu \,, \\
G^\mu &\equiv& \partial^\mu_x- j_0\left(\frac{1}{2}\hbar\Delta\right)F^{\mu\nu}\partial_\nu^p \,,\\
\label{Pi-mu}
\Pi^\mu &\equiv& p^\mu -\frac{1}{2}\hbar  j_1\left(\frac{1}{2}\hbar\Delta\right)F^{\mu\nu}\partial_\nu^p \,.
\end{eqnarray}
In the above, $j_{0}(z)$ and $j_{1}(z)$ are the zeroth and first order spherical Bessel functions, respectively, and
the triangle operator $\Delta\equiv \partial_p\cdot \partial_x$ denotes the mixed derivative.
Note that $\partial_{x}$ in the operator $\Delta$ acts only on $F^{\mu\nu}$ to its right but not on other functions.
Substituting the decomposition (\ref{decomposition}) into Eq. (\ref{eq-c}), we obtain coupled equations for components of the Wigner function
\begin{eqnarray}
\label{eq-F}
K^\mu\mathscr{V}_\mu -m\mathscr{F}&=&0,\\
\label{eq-P}
iK^\mu\mathscr{A}_\mu-m\mathscr{P}&=&0,\\
\label{eq-V}
K_\mu\mathscr{F}-iK^\nu\mathscr{S}_{\mu\nu}-m\mathscr{V}_\mu&=&0,\\
\label{eq-A}
iK_\mu\mathscr{P}+\frac{1}{2}\epsilon_{\mu\nu\rho\sigma}K^\nu\mathscr{S}^{\rho\sigma}+m\mathscr{A}_\mu&=&0,\\
\label{eq-S}
i\left(K_\mu \mathscr{V}_\nu-K_\nu \mathscr{V}_\mu\right)
+\epsilon_{\mu\nu\rho\sigma}K^\rho \mathscr{A}^\sigma +m \mathscr{S}_{\mu\nu}&=&0.
\end{eqnarray}
These equations can be further decomposed into real and imaginary parts.
The real parts of the above equations read
\begin{eqnarray}
\label{eq-F-re}
\Pi^\mu\mathscr{V}_\mu &=& m\mathscr{F},\\
\label{eq-P-re}
-\hbar G^\mu\mathscr{A}_\mu &=& 2m\mathscr{P},\\
\label{eq-V-re}
\Pi_\mu\mathscr{F}+\frac{1}{2}\hbar G^\nu\mathscr{S}_{\mu\nu} &=& m\mathscr{V}_\mu,\\
\label{eq-A-re}
\hbar G_\mu\mathscr{P}-\epsilon_{\mu\nu\rho\sigma}\Pi^\nu\mathscr{S}^{\rho\sigma}&=& 2m\mathscr{A}_\mu,\\
\label{eq-S-re}
\frac{1}{2}\hbar\left(G_\mu \mathscr{V}_\nu-G_\nu \mathscr{V}_\mu\right)
-\epsilon_{\mu\nu\rho\sigma}\Pi^\rho \mathscr{A}^\sigma &=& m \mathscr{S}_{\mu\nu}\,,
\end{eqnarray}
which are linearly proportional to the particle mass {on the right-hand side of the equations}, while the imaginary parts read
\begin{eqnarray}
\label{eq-F-im}
\hbar G^\mu\mathscr{V}_\mu &=&0,\\
\label{eq-P-im}
\Pi^\mu\mathscr{A}_\mu &=&0,\\
\label{eq-V-im}
\frac{1}{2}\hbar G_\mu\mathscr{F}-\Pi^\nu\mathscr{S}_{\mu\nu}&=&0,\\
\label{eq-A-im}
\Pi_\mu\mathscr{P}+\frac{1}{4}\hbar\epsilon_{\mu\nu\rho\sigma}G^\nu\mathscr{S}^{\rho\sigma} &=&0,\\
\label{eq-S-im}
\left(\Pi_\mu \mathscr{V}_\nu-\Pi_\nu \mathscr{V}_\mu\right)
+\frac{1}{2}\hbar\epsilon_{\mu\nu\rho\sigma}G^\rho \mathscr{A}^\sigma &=&0\,,
\end{eqnarray}
without explicit mass dependence.

We see that there are 16 components of
the Wigner function which satisfy above 32 equations.
In this sense quantum transport theory is much more complicated than classical one in which
only the phase space distribution $f(t,{\bf x},{\bf p})$ and the corresponding Vlasov-Boltzmann equation (\ref{Boltzmann-eq}) are involved.
Therefore it is important to disentangle the coupled equations for different components of the Wigner function as much as possible to reduce the number of independent variables and equations.
The key method is to use the semiclassical expansion in the Planck constant in which all operators and functions are expanded in powers of $\hbar$.
The expansions of the Wigner function $W$ and the operators $\Pi^{\mu}$ and $G^{\mu}$ are
\begin{equation}
W = \sum _{k=0}^\infty \hbar^{k}W^{(k)},\,\,\,
\Pi ^{\mu}=\sum _{k=0}^\infty \hbar^{2k} \Pi ^{(2k)\mu},\,\,\,
G^{\mu}=\sum _{k=0}^\infty \hbar^{2k} G^{(2k)\mu}\,,
\label{eq:expansion}
\end{equation}
where $k$ denote nonnegative integers, $W$ can be replaced of any of its components:
$\mathscr{F}$, $\mathscr{P}$, $\mathscr{V}_\mu$, $\mathscr{A}_\mu$ and $\mathscr{S}_{\mu\nu}$.
The expansion of $\Pi ^{\mu}$ and $G^{\mu}$ can be obtained by an expansion of Bessel functions and contains only even-order terms
\begin{eqnarray}
\Pi^{{(0)}\mu} &=& p^\mu,\\
\Pi^{{(2k)}\mu} & = & \frac{(-1)^{k}k}{2^{2k-1}(2k+1)!}
\Delta ^{2k-1} F^{\mu\nu}\partial_{\nu}^{p} \ \ \  (k\ge 1), \\
G^{{(2k)}\mu} & = & \frac{(-1)^{k+1}}{2^{2k}(2k+1)!}\Delta ^{2k}
F^{\mu\nu}\partial_{\nu}^{p}\ \ \    (k\ge 0).
\label{g-pi-n}
\end{eqnarray}
To $O(\hbar^2)$ these operator are
\begin{eqnarray}
\Pi^{\mu} & = &  p^{\mu}-\frac{1}{12}\hbar^2 \Delta F^{\mu\nu}\partial_\nu^p \,, \nonumber \\
G^{\mu} & = &   \partial_{x}^{\mu}- F^{\mu\nu}\partial_{\nu}^{p}+\frac{1}{24}\hbar^2
\Delta^2 F^{\mu\nu}\partial_\nu^p \,.
\label{op-2nd}
\end{eqnarray}
{We note that  $\hbar$ expansion of Bessel functions is equivalent to  expansion in $\Delta$ for the operators $\Pi^{\mu}$ and $G^{\mu}$, which implies that}
the condition for the validity of the expansion is
\begin{eqnarray}
\Delta R_F \Delta P_W \gg \hbar \,,
\end{eqnarray}
where $\Delta R_F$ denotes the spatial scale for the variation of the electromagnetic field $F^{\mu\nu}(x)$, while $\Delta P_W$ denotes the momentum scale for the variation of the Wigner function.


\section{Disentanglement of Wigner functions and equations}
\label{sec-disentangle}
The quantum transport theory in covariant Wigner functions are characterized by a set of coupled equations for their components. However one can show that the number of independent functions and the corresponding equations can be reduced in semiclassical expansion. In this section, we will show how to disentangle the components of Wigner functions and their equations in semiclassical expansion in powers of $\hbar$. We will first look at the case of massless fermions and then that of massive fermions.
{ The  results for massless fermions are mainly from Refs.~\citen{Gao:2018wmr} and \citen{Gao:2019zhk}
while those  for  massive fermions are from Refs.~\citen{Vasak:1987um} and \citen{Gao:2019znl}.}

\subsection{Massless fermions}
\label{subsec-massless}
For massless fermions with $m=0$, {the spin degree of freedom can be replaced by the chirality and}
the Dirac theory has a chiral symmetry, which is associated with a separate conservation of left- and right-handed fermion numbers at classical level. With chiral symmetry, the set of equations for components of Wigner functions are decoupled into two independent sets. In one set of equations, only $\mathscr{V}_\mu$ and $\mathscr{A}_\mu$ are involved,
\begin{eqnarray}
\label{eq-F-re-chiral}
\Pi^\mu\mathscr{V}_\mu &=& 0,\\
\label{eq-P-im-chiral}
\Pi^\mu\mathscr{A}_\mu &=&0,\\
\label{eq-F-im-chiral}
\hbar G^\mu\mathscr{V}_\mu &=&0,\\
\label{eq-P-re-chiral}
\hbar G^\mu\mathscr{A}_\mu &=& 0,\\
\label{eq-S-re-chiral}
\Pi_\mu \mathscr{V}_\nu- \Pi_\nu \mathscr{V}_\mu
+\frac{1}{2}\hbar \epsilon_{\mu\nu\rho\sigma}G^\rho \mathscr{A}^\sigma &=& 0,\\
\label{eq-S-im-chiral}
\Pi_\mu \mathscr{A}_\nu- \Pi_\nu \mathscr{A}_\mu
+\frac{1}{2}\hbar \epsilon_{\mu\nu\rho\sigma}G^\rho \mathscr{V}^\sigma &=& 0\,,
\end{eqnarray}
where the last equation can be obtained by contracting the antisymmetric tensor $\epsilon^{\mu\nu\alpha\beta}$ with Eq. (\ref{eq-S-re}).
It is obvious that the equations for $\mathscr{V}_\mu$ and $\mathscr{A}_\mu$ are symmetric for an interchange of $\mathscr{V}_\mu$ and $\mathscr{A}_\mu$ as a result of chiral symmetry.
In another set of equations,  $\mathscr{F}$, $\mathscr{P}$ and $\mathscr{S}_{\mu\nu}$ are involved
\begin{eqnarray}
\label{eq-V-re-chiral}
\Pi_\mu\mathscr{F}+\frac{1}{2}\hbar G^\nu\mathscr{S}_{\mu\nu} &=& 0,\\
\label{eq-A-re-chiral}
-\hbar G_\mu\mathscr{P}+\epsilon_{\mu\nu\rho\sigma}\Pi^\nu\mathscr{S}^{\rho\sigma}&=& 0,\\
\label{eq-V-im-chiral}
\frac{1}{2}\hbar G_\mu\mathscr{F}-\Pi^\nu\mathscr{S}_{\mu\nu}&=&0,\\
\label{eq-A-im-chiral}
\Pi_\mu\mathscr{P}+\frac{1}{4}\hbar\epsilon_{\mu\nu\rho\sigma}G^\nu\mathscr{S}^{\rho\sigma} &=&0\,.
\end{eqnarray}
Since we are mainly interested in the vector current $j^\mu$, the axial current $j_5^\mu$ and the energy-momentum tensor $T^{\mu\nu}$, which can be obtained from $\mathscr{V}_\mu$ and $\mathscr{A}_\mu$ by

\begin{eqnarray}
j^\mu = \int d^4 p \mathscr{V}^\mu,\ \ \
j_5^\mu = \int d^4 p \mathscr{A}^\mu,\ \ \
T^{\mu\nu}=\int d^4 p \mathscr{V}^\mu p^\nu \,,
\end{eqnarray}
we will focus on $\mathscr{V}_\mu$ and $\mathscr{A}_\mu$ at the chiral limit.
It is convenient to define chiral (left-hand and right-hand) Wigner functions from $\mathscr{V}_\mu$ and $\mathscr{A}_\mu$ as
\begin{equation}
\mathscr{J}^{\mu}_{s}=\frac{1}{2}\left(\mathscr{V}^{\mu}+s\mathscr{A}^{\mu}\right),
\label{vwfc}
\end{equation}
where $s=\pm$ is the chirality. It is obvious that the set of equations for chiral (left-hand and right-hand) Wigner functions are completely decoupled from each other
\begin{eqnarray}
\Pi^{\mu}\mathscr{J}_{s\mu}(x,p) & = & 0,\nonumber \\
G^{\mu}\mathscr{J}_{s\mu}(x,p) & = & 0,\nonumber \\
2s(\Pi^{\mu}\mathscr{J}_{s}^{\nu}-\Pi^{\nu}\mathscr{J}_{s}^{\mu}) & = & -\hbar\epsilon^{\mu\nu\rho\sigma}G_{\rho}\mathscr{J}_{s\sigma}\, .
\label{eq:wig-eq-1}
\end{eqnarray}

In Ref.~\citen{Gao:2018wmr}, by using semi-classical expansion as a tool, it has been shown that
different components of the covariant chiral Wigner function defined in Eq.~(\ref{eq:wig-eq-1}) can be disentangled from each other
and only one component of $\mathscr{J}^\mu_s$ is independent.
The conclusion is quite general and important in solving Wigner equations so that it is proposed to called it
a theorem of disentanglement for chiral Wigner function (DWF theorem). The theorem reads:
\textit{For  massless spin-$1/2$ ferimions, among the four components of the  chiral Wigner function $\mathscr{J}^\mu_s$ with $\mu =0,1,2,3$,
only one of them is independent, and all other three are determined by it.
With the on-shell condition, the independent component is determined by only one independent kinetic equation
while other equations are satisfied automatically.}


A detailed proof of the DWF theorem up to any order in the semiclassical expansion is given in Ref.~\citen{Gao:2018wmr},
and the independent kinetic equation for the independent component is given order by order in the expansion.
In this review, we will restrict ourselves to the zeroth and first order in $\hbar$ to illustrate the proof and the meaning of the theorem.
The generalization to higher orders is straightforward and we refer the reader to the original paper~\cite{Gao:2018wmr}.

Up to the first order in $\hbar$, the operators $G^\mu$ and $\Pi^\mu$ can be simplified as

\begin{eqnarray}
\label{nabla-mu}
G^\mu =\nabla^\mu \equiv \partial^\mu_x- F^{\mu\nu}\partial_\nu^p,\ \ \ \ \ \ \
\Pi^\mu = p^\mu.
\end{eqnarray}

As we have mentioned that there are four components of the chiral Wigner functions $\mathscr{J}_{s}^\mu(x,p)$.
We will show that there is only one independent component, the other three can be derived from it.
Note that we have the freedom to choose any one of the four components as the independent one.
To define an independent component, we introduce an auxiliary time-like four-vector $n^\mu$ normalized to $n^2=1$ \cite{Hidaka:2016yjf,Huang:2018wdl,Gao:2018wmr,Liu:2018xip}.
Then any vector $X^\mu$ can be decomposed into the component parallel or perpendicular to $n^\mu$,
\begin{equation}
X^\mu=X_n n^\mu + \bar X^\mu,
\label{decomp-n}
\end{equation}
with $X_n =n\cdot X$ and $\bar X \cdot n=0$.
We can regard $n^\mu $ as the four-velocity of an observer, then $\mathscr{J}_{sn}$ is the particle distribution function measured by him or her. In general, $n^\mu$ can depend on space-time coordinates and momentum. For simplicity, we assume $n^\mu$ is a constant vector except when we discuss about the CVE. With the decomposition (\ref{decomp-n}), the electromagnetic field tensor $F^{\mu\nu}$ can be put into the form
\begin{equation}
F^{\mu\nu}=E^\mu n^\nu -E^\nu n^\mu +\epsilon^{\mu\nu\rho\sigma}n_\rho B_\sigma\,.
\end{equation}
We see that the electric and magnetic fields $E^\mu=F^{\mu\nu}n_\nu$ and $B^\mu=\epsilon^{\mu\nu\rho\sigma}F_{\rho\sigma}/2$ are also defined with $n^\mu$.

With the decomposition (\ref{decomp-n}) and the semiclassical expansion, the set of equations for chiral Wigner functions at $O(\hbar^0)$ are
\begin{eqnarray}
\label{J-c1-1-0}
p_n \mathscr{J}_{sn}^{(0)}+{\bar p}\cdot \bar{\mathscr{J}}^{(0)}_s  &=&0,\\
\label{J-eq-1-0}
\nabla_n  \mathscr{J}_{sn}^{(0)}  +\bar \nabla\cdot  \bar{\mathscr{J}}^{(0)}_s&=& 0,\\
\label{J-c2-1a-0}
2s\left( \bar p_\mu \mathscr{J}_{sn}^{(0)} - p_n \bar{ \mathscr{J}}_{s\mu}^{(0)} \right)&=&0,\\
\label{J-c2-1b-0}
2s\left( \bar p_\mu \bar{\mathscr{J}}_{s\nu}^{(0)} - \bar p_\nu \bar{\mathscr{J}}_{s\mu}^{(0)}\right)&=&0 \,,
\end{eqnarray}
where quantities of $O(\hbar^0)$ are labelled by the superscript `(0)'.
It is natural to choose $\mathscr{J}_{sn}^{(0)}$ as the independent component, then
 it is easy to check that the space-like component $\bar{ \mathscr{J}}_{s\mu}^{(0)}$ is proportional to $\mathscr{J}_{sn}^{(0)}$ following from Eq. (\ref{J-c2-1a-0}) that
\begin{equation}
\label{bar-Jmu-0}
\bar{ \mathscr{J}}_{s\mu}^{(0)} = \bar p_\mu \frac{{\mathscr{J}}^{(0)}_{sn}}{  p_n }\,.
\end{equation}
With this relation, it is obvious that Eq. (\ref{J-c2-1b-0}) holds automatically and therefore is redundant.
Substituting (\ref{bar-Jmu-0}) into Eq. (\ref{J-c1-1-0}) yields
\begin{equation}
\label{Jn-0-onshell}
p^2 \frac{{\mathscr{J}}^{(0)}_{sn}}{ p_n } =0,
\end{equation}
which is the on-shell condition for free massless particle. The general solution of $\mathscr{J}_{sn}^{(0)}$ from Eq.~(\ref{Jn-0-onshell}) is given by
\begin{equation}
\label{Jn-0}
\frac{ {\mathscr{J}}^{(0)}_{sn}}{p_n} = f^{(0)}_s \delta\left(p^2\right)\,,
\end{equation}
where $ f^{(0)}_s$ is an arbitrary scalar function of $x$ and $p$ without singularity at $p^2=0$. From the solution (\ref{Jn-0}), $f^{(0)}_s$ can depend on $n^\mu$ in principle.
For higher order contribution $f^{(k)}_s\ (k\ge 1)$, it can be verified that $f^{(k)}_s$ indeed depend on $n^\mu$ but $f^{(0)}_s$ at the zeroth order is independent of $n^\mu$.
We will discuss about it at the end of this section.
Now let us combine Eq. (\ref{bar-Jmu-0}) with Eq. (\ref{Jn-0}), we obtain the chiral Wigner function in the vector form
\begin{equation}
\label{Jmu-0}
{\mathscr{J}}_{s\mu}^{(0)} = p_\mu f^{(0)}_s \delta\left(p^2\right).
\end{equation}
Substituting the above form into Eq. (\ref{J-eq-1-0}), we obtain the kinetic equation for $f^{(0)}_s$
\begin{equation}
\label{kinetic-eq-0}
\nabla_\mu \left[ p^\mu  f^{(0)}_s \delta \left(p^2\right)\right] =0 \,,
\end{equation}
which is just the covariant Vlasov equation in phase space. After an integration over $p_n$ with $n^\mu=(1,0,0,0)$, we can reproduce the usual Vlasov equation given in Eq. (\ref{Boltzmann-eq}) without the collision term.


Similar to the zeroth order, the set of equations for chiral Wigner functions at the first order read
\begin{eqnarray}
\label{J-c1-1-1}
p_n \mathscr{J}_{sn}^{(1)}+{\bar p}\cdot \bar{\mathscr{J}}^{(1)}_s  &=&0,\\
\label{J-eq-1-1}
\nabla_n  \mathscr{J}_{sn}^{(1)}  +\bar \nabla\cdot  \bar{\mathscr{J}}^{(1)}_s&=& 0,\\
\label{J-c2-1a-1}
 2s\left[ \bar p_\mu \mathscr{J}_{sn}^{(1)} - p_n \bar{ \mathscr{J}}_{s\mu}^{(1)} \right]
&=&- \epsilon_{\mu\nu\rho\sigma} n^\nu \bar \nabla^\rho\bar{ \mathscr{J}}^{(0)\sigma}_s ,\\
\label{J-c2-1b-1}
2s\left( \bar p_\mu \bar{\mathscr{J}}_{s\nu}^{(1)} - \bar p_\nu \bar{\mathscr{J}}_{s\mu}^{(1)}\right)
&=&-\epsilon_{\mu\nu\rho\sigma} n^\rho  \left[\nabla_n \bar{ \mathscr{J}}^{(0)\sigma}_s - \bar\nabla^\sigma \mathscr{J}_{sn}^{(0)} )  \right] \,.
\end{eqnarray}
From Eq. (\ref{J-c2-1a-1}), we can express $\bar{\mathscr{J}}^{(1)}_s $ as a function of $\mathscr{J}_{sn}^{(1)}$ and $\mathscr{J}_{sn}^{(0)}$
\begin{equation}
\label{bar-Jmu-1}
\bar{\mathscr{J}}_{s\mu}^{(1)} = \bar p_\mu \frac{{\mathscr{J}}^{(1)}_{sn}}{ p_n }
+\frac{s}{2p_n}\epsilon^{\mu\nu\rho\sigma} n_\nu \bar \nabla_\rho\bar{ \mathscr{J}}_{s\sigma}^{(0)}\,,
\end{equation}
where $\bar{ \mathscr{J}}_{s\sigma}^{(0)}$ is given by Eq. (\ref{bar-Jmu-0}).
Inserting $\bar{\mathscr{J}}_{s\mu}^{(1)}$ into Eq. (\ref{J-c1-1-1}), we obtain
\begin{equation}
p^2 \frac{{\mathscr{J}}^{(1)}_{sn}}{p_n} = \frac{s B\cdot p }{p_n} \frac{{\mathscr{J}}^{(0)}_{sn}}{p_n}=\frac{s B\cdot p }{p_n} f^{(0)}_s\delta(p^2)\,.
\end{equation}
The general solution to ${\mathscr{J}}^{(1)}_{sn}$ is then
\begin{equation}
\label{Jn-1}
{\mathscr{J}}_{sn}^{(1)} = p_n f^{(1)}_s \delta\left(p^2\right)
- s B\cdot p f^{(0)}_s  \delta'\left(p^2\right)\,,
\end{equation}
where we have introduced the first order distribution $f^{(1)}_s $, and the second term provides a correction to the on-shell condition of free fermions from the magnetic field which is a quantum effect.
Combining Eq. (\ref{Jn-1}) with Eq. (\ref{bar-Jmu-1}) leads to the full form of the chiral Wigner function
\begin{eqnarray}
\label{Jmu-1}
{\mathscr{J}}_{s\mu}^{(1)} &=& p_\mu f^{(1)}_s \delta\left(p^2\right)
- p_\mu \frac{s B\cdot p }{ p_n } f^{(0)}_s  \delta'\left(p^2\right)\nonumber\\
&& - \frac{s}{2p_n}\epsilon_{\mu\nu\rho\sigma} n^\nu \nabla^\sigma
 \left[ p^\rho f^{(0)}_s \delta\left(p^2\right)\right]\,.
\end{eqnarray}
If Eq. (\ref{bar-Jmu-1}) is inserted into Eq. (\ref{J-c2-1b-1}),
it is obvious that the $f^{(1)}_s$ term does not contribute and only the terms with $f^{(0)}_s$ survive.
We can verify that Eq. (\ref{J-c2-1b-1}) holds automatically with Eqs. (\ref{J-c1-1-0}-\ref{J-c2-1a-0}) or equivalently Eqs. (\ref{bar-Jmu-0},\ref{Jn-0}) being fulfilled.
{ In this way, we have shown DWF theorem to the first order in the semi-classical expansion.}
Such a procedure can be taken for equations of higher orders, we can show that the counterpart of Eq. (\ref{J-c2-1b-1})
at the $k$-th order does not depend on $f^{(k)}_s$ and holds automatically when $(k-1)$-th order equations are satisfied.
This iterative process have been demonstrated by mathematical induction leading to the DWF theorem in Ref.~\citen{Gao:2018wmr}.
Substituting Eq. (\ref{Jmu-1}) into Eq. (\ref{J-eq-1-1}) gives a kinetic equation for the first order distribution.

The total chiral Wigner function ${\mathscr{J}}_{sn}$ to the first order is given by the sum of Eq. (\ref{Jn-0}) and (\ref{Jn-1})
\begin{eqnarray}
\label{Jn-sum}
{\mathscr{J}}_{sn} = {\mathscr{J}}_{sn}^{(0)} + \hbar {\mathscr{J}}_{sn}^{(1)}
&=&p_n\left(f^{(0)}_s + f^{(1)}_s \right)\delta\left(p^2\right)
- s \hbar B\cdot p  f^{(0)}_s  \delta'\left(p^2\right)\nonumber\\
&\approx& p_n f_s \delta \left(p^2-\frac{s \hbar B\cdot p }{p_n}\right)\,,
\end{eqnarray}
where $f_s\equiv f^{(0)}_s+f^{(1)}_s$.
We can see that the quantum correction to the on-shell condition arises from the magnetic moment energy of massless fermions that has been absorbed into the on-shell $\delta$ function.
Taking a sum of Eq. (\ref{Jmu-0}) and (\ref{Jmu-1}), we obtain the first order chiral Wigner function
\begin{eqnarray}
\label{Jmu-sum}
{\mathscr{J}}_{s\mu} &=& {\mathscr{J}}_{s\mu}^{(0)} + \hbar {\mathscr{J}}_{s\mu}^{(1)}\nonumber\\
&=&p_\mu \left(f^{(0)}_s + f^{(1)}_s \right)\delta\left(p^2\right)
- p_\mu \frac{s B\cdot p }{ p_n } f^{(0)}_s  \delta'\left(p^2\right) \nonumber\\
&& - \frac{s}{2p_n}\epsilon_{\mu\nu\rho\sigma} n^\nu    \nabla^\sigma \left[ p^\rho f^{(0)}_s \delta\left(p^2\right)\right]\nonumber\\
&\approx& {\left( g_{\mu\nu}  + \frac{s\hbar}{2p_n}\epsilon_{\mu\nu \rho \sigma} n^\rho    \nabla^\sigma \right)
\left[ p^\nu f_s \delta \left(p^2-\frac{s \hbar B\cdot p }{p_n}\right)\right]}\,.
\end{eqnarray}
Then the covariant chiral kinetic equation is given by a sum of Eq. (\ref{J-eq-1-0}) and (\ref{J-eq-1-1})
\begin{eqnarray}
\label{ccke}
\nabla_\mu \left\{{\left( g^{\mu\nu}  + \frac{s\hbar}{2p_n}\epsilon^{\mu\nu \rho \sigma} n_\rho    \nabla_\sigma \right)
\left[ p_\nu f_s \delta \left(p^2 - \frac{s \hbar B\cdot p }{p_n}\right)\right]} \right\}=0 \,.
\end{eqnarray}

To obtain the Vlasov equation as in Eq. (\ref{Boltzmann-eq}) in three-momentum space, we set $n^\mu=(1,0,0,0)$ and perform an integration over $p_{0}$. The positive $p_{0}$ part can be extracted by setting the integral range to $(0,\infty)$ which gives the chiral kinetic equation for particles, while the negative $p_{0}$ part corresponds to the range $(-\infty,0)$ which gives the equation for antiparticles.
After completing an integration over $p_0$ from $0$ to $+\infty$,
we obtain the chiral kinetic equation for particles with helicity $s$
\begin{eqnarray}
\label{cke-particle}
 \left(1 + s\hbar {\bf B}\cdot{\pmb \Omega}_p\right)  \partial_t f_s(t,{\bf x},{\bf p})& &\nonumber\\
+ \left[{\pmb v} + s\hbar ( \hat{\bf p} \cdot {\pmb \Omega}_p ) {\bf B}
+ {s\hbar}{\bf E}\times{\pmb \Omega}_p\right] \cdot {\pmb \nabla}_x f_s(t,{\bf x},{\bf p}) & &\nonumber\\
+\left[\tilde {\bf E} + {\pmb v}\times {\bf B}
 + {s\hbar} ({\bf E}\cdot {\bf B} )  {\pmb \Omega}_p \right]\cdot {\pmb \nabla}_p   f_s(t,{\bf x},{\bf p}) & &  \nonumber\\
 + {s\hbar}{\bf E}\cdot{\bf B} \left(  {\pmb \nabla}_p\cdot {\pmb \Omega}_p\right) f_s(t,{\bf x},{\bf p}) &=& 0 \,, \hspace{1cm}
\end{eqnarray}
where $\hat{\bf p}={\bf p}/{|\bf p|}$ denotes the direction of a three-momentum (unit vector),
${\pmb \Omega}_p = {\bf p}/(2|{\bf p}|^3)$ is the Berry curvature in momentum space,
and other symbols are defined as
\begin{eqnarray}
 f_s(t,{\bf x},{\bf p}) &=& \left.f_s(x,p)\right|_{p_0=E_p^+}\,, \\
 E_p^+ &=& |{\bf p}|(1-\hbar s {\bf B}\cdot {\pmb \Omega}_p )\,,\\
{\pmb v} &=&  {\pmb \nabla}_p E_p^+ \,, \\
\tilde{\bf E} &=& {\bf E} - {\pmb \nabla}_x E_p^+ \,.
\end{eqnarray}
By an integration over $p_0$ from  $-\infty$ to $0$ and replacing ${\bf p}$ and $s$ with $-{\bf p}$ and $-s$ respectively, we obtain the chiral kinetic equation for antiparticles
with helicity $s$
\begin{eqnarray}
\label{cke-antiparticle}
 \left(1 - s\hbar {\bf B}\cdot{\pmb \Omega}_p\right)  \partial_t\bar f_s^{\,\textrm{t}}(t,{\bf x},{\bf p})& &\nonumber\\
- \left[{\pmb v} + s\hbar ( \hat{\bf p} \cdot {\pmb \Omega}_p ) {\bf B}
+ {s\hbar}{\bf E}\times{\pmb \Omega}_p\right] \cdot {\pmb \nabla}_x \bar f_s^{\,\textrm{t}}(t,{\bf x},{\bf p}) & &\nonumber\\
-\left[\tilde {\bf E} + {\pmb v}\times {\bf B}
 - {s\hbar}( {\bf E}\cdot {\bf B} )  {\pmb \Omega}_p \right]\cdot {\pmb \nabla}_p  \bar f_s^{\,\textrm{t}}(t,{\bf x},{\bf p}) & &  \nonumber\\
 + {s\hbar}{\bf E}\cdot{\bf B} \left(  {\pmb \nabla}_p\cdot {\pmb \Omega}_p\right)\bar f_s^{\,\textrm{t}}(t,{\bf x},{\bf p}) &=&0, \hspace{1cm}
\end{eqnarray}
where
\begin{eqnarray}
\bar f_s^{\,\textrm{t}}(t,{\bf x},{\bf p}) &=& \left.f_s(x,p)\right|_{p_0=E_p^-}\,, \\
 E_p^- & =&- |{\bf p}|(1+\hbar s {\bf B}\cdot {\pmb \Omega}_p )\,, \\
{\pmb v} &=&  {\pmb \nabla}_p E_p^- \,,\\
\tilde{\bf E} &=&{\bf E} - {\pmb \nabla}_x E_p^-\,.
\end{eqnarray}
The superscript `t' in $\bar f_s^{\textrm{t}}(t,{\bf x},{\bf p})$ denotes the total distribution
defined as a sum of the normal distribution $\bar f_s(t,{\bf x},{\bf p})$ and the vacuum contribution
$\bar f_s^{\,\textrm v}(t,{\bf x},{\bf p})$,
\begin{eqnarray}
\bar f_s^{\,\textrm{t}}(t,{\bf x},{\bf p}) = \bar f_s(t,{\bf x},{\bf p}) + \bar f_s^{\textrm v}(t,{\bf x},{\bf p})\,.
\end{eqnarray}
The vacuum contribution originates from the definition (\ref{wigner}) of the Wigner function without normal ordering. There is no vacuum contribution in the particle distribution $ f_s(t,{\bf x},{\bf p})$.
In the free case we have $\bar f_s^{\,\textrm{v}}=-1$, the kinetic equation for the normal distribution
$\bar f_s(t,{\bf x},{\bf p})$ reads
\begin{eqnarray}
\label{cke-antiparticle-normal}
 \left(1 - s\hbar {\bf B}\cdot{\pmb \Omega}_p\right)  \partial_t\bar f_s(t,{\bf x},{\bf p})& &\nonumber\\
- \left[{\pmb v} + s\hbar ( \hat{\bf p} \cdot {\pmb \Omega}_p ) {\bf B}
+ {s\hbar}{\bf E}\times{\pmb \Omega}_p\right] \cdot {\pmb \nabla}_x \bar f_s(t,{\bf x},{\bf p}) & &\nonumber\\
-\left[\tilde {\bf E} + {\pmb v}\times {\bf B}
 - {s\hbar} ({\bf E}\cdot {\bf B})   {\pmb \Omega}_p \right]\cdot {\pmb \nabla}_p  \bar f_s(t,{\bf x},{\bf p}) & &  \nonumber\\
 + {s\hbar}{\bf E}\cdot{\bf B} \left(  {\pmb \nabla}_p\cdot {\pmb \Omega}_p\right)\left[\bar f_s(t,{\bf x},{\bf p})-1\right] &=&0 \,. \hspace{1cm}
\end{eqnarray}
The inhomogeneous term independent of $\bar f_s(t,{\bf x},{\bf p})$ comes from vacuum or Dirac sea
which plays a central role in deriving chiral anomaly.
We note that we have taken the vacuum contribution $\bar f^{\,\textrm{v}}=-1$ for free antifermions in Eq. (\ref{cke-antiparticle}),
{ but in principle there are possible quantum corrections to $\bar f^{\,\textrm{v}}$ at $O(\hbar)$. We assume these higher order contributions   remain in  $\bar f_s(t,{\bf x},{\bf p})$.}

In classical transport theory, the particle distribution function $f(t,{\bf x},{\bf p})$ in Eq. (\ref{Boltzmann-eq}) transforms as a scalar function when changing the reference frame. Now let us discuss whether this conclusion still holds in quantum transport theory at chiral limit. As we mentioned above,
we can regard the auxiliary vector $n^\mu $ as the observer's four-velocity,
and then ${\mathscr{J}}_{sn}$ measures the particle distribution function in a general Lorentz frame corresponding to $n^\mu$. Certainly, we can choose another vector $n^{\prime}_\mu$ to make the decomposition (\ref{decomp-n}).
Then ${\mathscr{J}}^{(0)\mu}_s$ and ${\mathscr{J}}^{(1)\mu}_s$ in Eqs. (\ref{Jmu-0},\ref{Jmu-1})
are expressed in terms of $n^{\prime}_\mu$
\begin{eqnarray}
{\mathscr{J}}^{(0)\mu}_s&=& p^\mu \frac{{\mathscr{J}}^{(0)}_{sn'}}{ p_{n'}},\\
{\mathscr{J}}^{(1)\mu}_s &=& p^\mu \frac{{\mathscr{J}}^{(1)}_{sn'}}{ p_{n'}}
 + \frac{s}{2 p_{n'}}\epsilon^{\mu\nu\rho\sigma} n' _\nu    \nabla_\rho {\mathscr{J}}^{(0)}_{s\sigma}\,,
\end{eqnarray}
which have a different but equivalent form to Eqs. (\ref{Jmu-0},\ref{Jmu-1}).
On the other hand, the Wigner function ${\mathscr{J}}_\mu^{(0)}$ and ${\mathscr{J}}_\mu^{(1)}$ must be independent of the frame we choose, so these two decompositions must be identical to each other.
This leads to the transformation rule for distribution functions
\begin{eqnarray}
\delta \left(\frac{{\mathscr{J}}^{(0)}_{sn} }{p_n}\right)
&\equiv& \frac{{\mathscr{J}}^{(0)}_{sn'} }{p_{n'}}-\frac{{\mathscr{J}}^{(0)}_{sn} }{p_n}=0,\\
\delta \left(\frac{{\mathscr{J}}^{(1)}_{sn} }{p_n}\right)
&\equiv& \frac{{\mathscr{J}}^{(1)}_{sn'} }{p_{n'}}-\frac{{\mathscr{J}}^{(1)}_{sn} }{p_n}
=- \frac{ s \epsilon^{\lambda\nu\rho\sigma} n_\lambda n^{\prime}_\nu
\nabla_\rho {\mathscr{J}}^{(0)}_{s\sigma} } {2 \left(n^\prime \cdot p\right)\left( n\cdot p\right) },
\end{eqnarray}
or equivalently the transformation rule for $f^{(0)}_s$ and $f^{(1)}_s$,
\begin{eqnarray}
\label{df0}
\delta\left(p^2\right) \delta  f^{(0)}_s &=& 0 \,, \nonumber\\
\label{df1}
\delta\left(p^2\right)\delta f^{(1)}_s
&=&\left(\frac{n'_\nu \tilde F^{\nu\lambda}p_\lambda}{n'\cdot p}
-\frac{n_\nu \tilde F^{\nu\lambda}p_\lambda}{n\cdot p}\right)s\delta'\left(p^2\right) f^{(0)}_s\nonumber\\
& &-\frac{ s \epsilon^{\lambda\nu\rho\sigma} n_\lambda n'_\nu  } {2 \left(n'\cdot p\right)\left( n\cdot p\right) }
\nabla_\rho\left[ p_\sigma f^{(0)}_s \delta\left(p^2\right) \right] \,,
\end{eqnarray}
where we have defined the dual field strength tensor
$\tilde F^{\nu\lambda}=\epsilon^{\nu\lambda\alpha\beta}F_{\alpha\beta}/2$.
We see that the zeroth-order distribution function $f^{(0)}_s$ is a Lorentz scalar and does not depend on the observer's frame, consistent with the classical picture. However the quantum correction modifies this conclusion: the non-trivial transformation rule at the first order is related to the side-jump term first proposed in a study of Lorentz invariance of chiral kinetic theory \cite{Chen:2014cla} and later verified
in the  Wigner funciton method from quantum field theory \cite{Hidaka:2016yjf,Huang:2018wdl,Gao:2018wmr}.
Therefore the distribution function and its transformation in different frame can be defined by the Wigner function in a transparent way.


\subsection{Massive fermions}
\label{subsec-massive}
Quantum kinetic equations for massive fermions in terms of Wigner functions are much more complicated than massless fermions because all components of Wigner functions are entangled in these equations.
However, with the help of semiclassical expansion, we can reduce the number of independent equations {substantially.}
{We can choose the independent components of Wigner functions in several different ways. We will follow the procedure  given in Refs.~\citen{Vasak:1987um,Gao:2019znl}.
Other possible choices can be found in Refs.~\citen{Weickgenannt:2019dks,Hattori:2019ahi,Wang:2019moi}}
To this end, we expand all functions and operators in $\hbar$ as in Eq. (\ref{eq:expansion}).
At the zeroth order in $\hbar$ the equations for Wigner function components read
\begin{eqnarray}
\label{eq-F-re-hbar-0}
p^\mu\mathscr{V}_\mu^{(0)} &=& m\mathscr{F}^{(0)},\\
\label{eq-P-re-hbar-0}
0 &=& m\mathscr{P}^{(0)},\\
\label{eq-V-re-hbar-0}
p_\mu\mathscr{F}^{(0)}&=& m\mathscr{V}_\mu^{(0)},\\
\label{eq-A-re-hbar-0}
-\frac{1}{2}\epsilon_{\mu\nu\rho\sigma}p^\nu\mathscr{S}^{(0)\rho\sigma}&=& m\mathscr{A}_\mu^{(0)},\\
\label{eq-S-re-hbar-0}
-\epsilon_{\mu\nu\rho\sigma}p^\rho \mathscr{A}^{(0)\sigma} &=& m \mathscr{S}_{\mu\nu}^{(0)}\,,
\end{eqnarray}
and
\begin{eqnarray}
\label{eq-F-im-hbar-0}
 \nabla^\mu\mathscr{V}_\mu^{(0)} &=&0,\\
\label{eq-P-im-hbar-0}
p^\mu\mathscr{A}_\mu^{(0)} &=&0,\\
\label{eq-V-im-hbar-0}
p^\nu\mathscr{S}_{\mu\nu}^{(0)}&=&0,\\
\label{eq-A-im-hbar-0}
p_\mu\mathscr{P}^{(0)} &=&0,\\
\label{eq-S-im-hbar-0}
p_\mu \mathscr{V}_\nu^{(0)}-p_\nu \mathscr{V}_\mu^{(0)} &=&0\,.
\end{eqnarray}
It is convenient to choose $\mathscr{F}^{(0)}$ and $\mathscr{A}^{(0)}_\mu$ as independent components of Wigner functions, from which all other components can be derived. From Eqs. (\ref{eq-P-re-hbar-0},\ref{eq-V-re-hbar-0},\ref{eq-S-re-hbar-0}) we obtain
\begin{eqnarray}
\label{P-0-a0}
\mathscr{P}^{(0)} & = & 0 \,, \\
\label{V-0-a0}
\mathscr{V}_{\mu}^{(0)} &=& \frac{1}{m}p_\mu  \mathscr{F}^{(0)}\,, \\
\label{S-0-a0}
\mathscr{S}_{\mu\nu}^{(0)} &=&-\frac{1}{m} \epsilon_{\mu\nu\rho\sigma}p^\rho \mathscr{A}^{(0)\sigma}\,.
\end{eqnarray}
It is obvious that Eqs. (\ref{eq-V-im-hbar-0}-\ref{eq-S-im-hbar-0})
are fulfilled automatically by Eqs. (\ref{P-0-a0}-\ref{S-0-a0}).
Substituting Eqs. (\ref{V-0-a0},\ref{S-0-a0}) into  Eqs. (\ref{eq-F-re-hbar-0},\ref{eq-A-re-hbar-0}) leads to
on-shell conditions for $\mathscr{F}^{(0)}$ and $\mathscr{A}_{\mu}^{(0)}$, respectively,
\begin{eqnarray}
\label{F-0-a}
(p^2-m^2)\mathscr{F}^{(0)} &=& 0,\\
(p^2-m^2)\mathscr{A}_{\mu}^{(0)} &=& 0\,,
\end{eqnarray}
which indicates that both $\mathscr{F}^{(0)}$ and $\mathscr{A}_{\mu}^{(0)}$ take the following forms
\begin{eqnarray}
\label{F-0-b}
\mathscr{F}^{(0)} &=&\delta\left(p^2-m^2\right)\mathcal{F}^{(0)} , \\
\label{A-0-b}
\mathscr{A}_{\mu}^{(0)} &=&\delta\left(p^2-m^2\right) \mathcal{A}_{\mu}^{(0)}\,.
\end{eqnarray}
Here an arbitrary scalar function $\mathcal{F}^{(0)}$ and axial vector function $\mathcal{A}_{\mu}^{(0)}$ are both nonsingular at $p^2-m^2=0$ and can only be determined by kinetic equations.
Substituting Eq. (\ref{V-0-a0}) into Eq. (\ref{eq-F-im-hbar-0}),
we obtain the kinetic equation for $\mathscr{F}^{(0)}$
\begin{eqnarray}
\label{Veq-0-a}
p^\mu \nabla_\mu\mathscr{F}^{(0)} &=&0 \,.
\end{eqnarray}
At the zeroth order in $\hbar$ there is a constraint equation (\ref{eq-P-im-hbar-0}) for $\mathscr{A}_{\mu}^{(0)}$. However, the kinetic equation for $\mathscr{A}_{\mu}^{(0)}$ will not show up until at the first order.


The equations for Wigner function's components at the first order read
\begin{eqnarray}
\label{eq-F-re-hbar-1}
p^\mu\mathscr{V}_\mu^{(1)} &=& m\mathscr{F}^{(1)},\\
\label{eq-P-re-hbar-1}
-\frac{1}{2} \nabla^\mu\mathscr{A}_\mu^{(0)} &=& m\mathscr{P}^{(1)},\\
\label{eq-V-re-hbar-1}
p_\mu\mathscr{F}^{(1)}+\frac{1}{2} \nabla^\nu\mathscr{S}_{\mu\nu}^{(0)} &=& m\mathscr{V}_\mu^{(1)},\\
\label{eq-A-re-hbar-1}
\frac{1}{2}\nabla_\mu\mathscr{P}^{(0)}-\frac{1}{2}\epsilon_{\mu\nu\rho\sigma}p^\nu\mathscr{S}^{(1)\rho\sigma}
&=& m\mathscr{A}_\mu^{(1)},\\
\label{eq-S-re-hbar-1}
\frac{1}{2}\left(\nabla_\mu \mathscr{V}_\nu^{(0)}-\nabla_\nu \mathscr{V}_\mu^{(0)}\right)
-\epsilon_{\mu\nu\rho\sigma}p^\rho \mathscr{A}^{(1)\sigma} &=& m \mathscr{S}_{\mu\nu}^{(1)},
\end{eqnarray}
and
\begin{eqnarray}
\label{eq-F-im-hbar-1}
 \nabla^\mu\mathscr{V}_\mu^{(1)} &=&0,\\
\label{eq-P-im-hbar-1}
p^\mu\mathscr{A}_\mu^{(1)} &=&0,\\
\label{eq-V-im-hbar-1}
\frac{1}{2} \nabla_\mu\mathscr{F}^{(0)}-p^\nu\mathscr{S}_{\mu\nu}^{(1)}&=&0,\\
\label{eq-A-im-hbar-1}
p_\mu\mathscr{P}^{(1)}+\frac{1}{4}\epsilon_{\mu\nu\rho\sigma}\nabla^\nu\mathscr{S}^{(0)\rho\sigma} &=&0,\\
\label{eq-S-im-hbar-1}
\left(p_\mu \mathscr{V}_\nu^{(1)}-p_\nu \mathscr{V}_\mu^{(1)}\right)
+\frac{1}{2}\epsilon_{\mu\nu\rho\sigma}\nabla^\rho \mathscr{A}^{(0)\sigma} &=&0.
\end{eqnarray}
From Eqs. (\ref{eq-P-re-hbar-1},\ref{eq-V-re-hbar-1},\ref{eq-S-re-hbar-1}), we can express
$\mathscr{P}^{(1)}$, $\mathscr{V}_\mu^{(1)}$ and $\mathscr{S}_{\mu\nu}^{(1)}$ as the functions of
$\mathscr{F}^{(1)}$ and  $\mathscr{A}_\mu^{(1)}$,
\begin{eqnarray}
\label{P-1-a}
\mathscr{P}^{(1)} &=&\frac{1}{2m} \nabla^\mu \mathscr{A}_{\mu}^{(0)},\\
\label{V-1-a}
\mathscr{V}_{\mu}^{(1)} &=& \frac{1}{m}p_\mu  \mathscr{F}^{(1)} - \frac{1}{2m^2}\epsilon_{\mu\nu\rho\sigma} \nabla^\nu\left( p^\rho \mathscr{A}^{(0)\sigma}\right),\\
\label{S-1-a}
\mathscr{S}_{\mu\nu}^{(1)} &=&-\frac{1}{m} \epsilon_{\mu\nu\rho\sigma}p^\rho \mathscr{A}^{(1)\sigma}
+\frac{1}{2m^2}\left[ \nabla_\mu( p_\nu\mathscr{F}^{(0)}) -  \nabla_\nu (p_\mu \mathscr{F}^{(0)}) \right]\,.
\end{eqnarray}
Substituting Eq. (\ref{V-1-a}) into Eq. (\ref{eq-F-re-hbar-1})
and Eq. (\ref{S-1-a}) with Eq. (\ref{P-0-a0}) into Eq. (\ref{eq-A-re-hbar-1}),
we obtain the modified on-shell condition for $\mathscr{F}^{(1)}$ and $\mathscr{A}_{\mu}^{(1)}$, respectively,
\begin{eqnarray}
\label{F-1-a}
(p^2-m^2)\mathscr{F}^{(1)} &=& -\frac{1}{m} p^\mu \tilde F_{\mu\nu}\mathscr{A}^{(0)\nu},\\
\label{A-1-a}
(p^2-m^2)\mathscr{A}_{\mu}^{(1)} &=& - \frac{1}{m} p^\nu \tilde F_{\mu\nu}\mathscr{F}^{(0)},
\end{eqnarray}
which implies the general form of $\mathscr{F}^{(1)}$ and $\mathscr{A}_{\mu}^{(1)}$ as
\begin{eqnarray}
\label{F-1-b}
\mathscr{F}^{(1)} &=&\delta\left(p^2-m^2\right) \mathcal{F}^{(1)}+\frac{1}{m} \tilde F_{\mu\nu} p^\mu \mathcal{A}^{(0)\nu}\delta'\left(p^2-m^2\right) , \\
\label{A-1-b}
\mathscr{A}_{\mu}^{(1)} &=& \delta\left(p^2-m^2\right)  \mathcal{A}_{\mu}^{(1)}  + \frac{1}{m}\epsilon_{\mu\nu\rho\sigma}p^\nu \tilde F_{\mu\nu }\mathcal{F}^{(0)}\delta'\left(p^2-m^2\right)\,,
\end{eqnarray}
where we have introduced the functions $\mathcal{F}^{(1)}$ and $\mathcal{A}_{\mu}^{(1)}$ as the first-order correction to $\mathcal{F}^{(0)}$ and $\mathcal{A}_{\mu}^{(0)}$.
It is straightforward to verify that Eqs. (\ref{eq-V-im-hbar-1}) and (\ref{eq-S-im-hbar-1}) are fulfilled automatically.
Substituting Eq. (\ref{V-1-a}) into Eq. (\ref{eq-F-im-hbar-1}), we obtain the kinetic equation for $\mathscr{F}^{(1)}$
\begin{equation}
\label{Veq-1-a}
p^\mu \nabla_\mu\mathscr{F}^{(1)} = \frac{1}{2m} p^\mu \Delta \left(\tilde F_{\mu\nu} \mathscr{A}^{\nu}_{(0)}\right),
\end{equation}
Substituting Eq. (\ref{P-1-a}) and Eq. (\ref{S-0-a0}) into Eq. (\ref{eq-A-im-hbar-1}) leads to the kinetic equation for
$\mathscr{A}_{\mu}^{(0)}$,
\begin{equation}
\label{PS-1-a1}
p^\nu \nabla_\nu  \mathscr{A}_{\mu}^{(0)} = F_{\mu\nu} \mathscr{A}^{(0)\nu}\,.
\end{equation}
In order to obtain the kinetic equation for $\mathscr{A}_{\mu}^{(1)}$, we need the second-order counterpart of Eq. (\ref{eq-A-im-hbar-1}). The procedure is similar to obtain Eq. (\ref{PS-1-a1}) and the result is
\begin{eqnarray}
\label{PS-2-a1}
p^\nu \nabla_\nu  \mathscr{A}_{\mu}^{(1)} &=&F_{\mu\nu} \mathscr{A}^{(1)\nu}
+\frac{1}{2m} p^\nu \Delta \tilde F^{\mu\nu}\mathscr{F}^{(0)}\,.
\end{eqnarray}
The constraint condition (\ref{eq-P-im-hbar-1}) has to be fulfilled by $\mathscr{A}_{\mu}^{(1)}$.
Just like the massless case, we can combine the zeroth and first order contributions
and put all components into compact forms
\begin{eqnarray}
\mathscr{F} & \equiv &\mathscr{F}^{(0)} +\hbar\mathscr{F}^{(1)}  \,, \\
\mathscr{A}_{\mu} & \equiv & \mathscr{A}_{\mu}^{(0)}+\hbar\mathscr{A}_{\mu}^{(1)}\,, \\
\label{P}
\mathscr{P} &\equiv& \mathscr{P}^{(0)} +\hbar \mathscr{P}^{(1)} = -\frac{\hbar}{2m} \nabla^\mu \mathscr{A}_{\mu}\,,\\
\label{Vmu}
\mathscr{V}_{\mu} &\equiv & \mathscr{V}_{\mu}^{(0)}+\hbar \mathscr{V}_{\mu}^{(1)}= \frac{1}{m}p_\mu \mathscr{F}
- \frac{\hbar}{2m^2}\epsilon_{\mu\nu\rho\sigma}\nabla^\nu (p^\rho \mathscr{A}^{\sigma})\,,\\
\label{Smunu}
\mathscr{S}_{\mu\nu}&\equiv& \mathscr{S}_{\mu\nu}^{(0)}+\hbar \mathscr{S}_{\mu\nu}^{(1)}
=-\frac{1}{m} \epsilon_{\mu\nu\rho\sigma}p^\rho \mathscr{A}^{\sigma}
+\frac{\hbar}{2m^2}\left[ \nabla_\mu  (p_\nu \mathscr{F}) -  \nabla_\nu ( p_\mu \mathscr{F}) \right].
\end{eqnarray}
The onshell conditions for $\mathscr{F}$ and $\mathcal{A}$ read
\begin{eqnarray}
\label{F-1-b-sum}
\mathscr{F} &=&\delta\left(p^2-m^2\right) \mathcal{F}+\frac{\hbar}{m} \tilde F_{\mu\nu} p^\mu \mathcal{A}^{\nu}\delta'\left(p^2-m^2\right)\,, \\
\label{A-1-b-sum}
\mathscr{A}_{\mu}&=& \delta\left(p^2-m^2\right)  \mathcal{A}_{\mu}  +\frac{\hbar}{m} p^\nu \tilde F_{\mu\nu}\mathcal{F}\delta'\left(p^2-m^2\right)\,,
\end{eqnarray}
The covariant kinetic equations read
\begin{eqnarray}
\label{F-eq-4v}
& & p\cdot\nabla 
\left[\mathcal{F}\delta\left(p^2-m^2\right)
+\frac{\hbar}{m}\tilde  F_{\mu\nu} p^\mu\mathcal{A}^{\nu}\delta'\left(p^2-m^2\right)\right]\nonumber\\
&=&\frac{\hbar}{2m}(\partial_\lambda^x \tilde F_{\mu\nu})\partial^\lambda_p\left[p^\mu \mathcal{A}^{\nu}\delta\left(p^2-m^2\right)\right], \\
\label{A-eq-4v}
& & p\cdot\nabla
\left[ \mathcal{A}_{\mu}\delta\left(p^2-m^2\right)
+\frac{\hbar}{m}  p^\nu \tilde F_{\mu\nu}\mathcal{F}\delta'\left(p^2-m^2\right)  \right]\nonumber\\
&=&  F_{\mu\nu}\left[ \mathcal{A}^{\nu}\delta\left(p^2-m^2\right)
+\frac{\hbar}{m} p_\lambda \tilde F^{\nu\lambda}\mathcal{F}\delta'\left(p^2-m^2\right)\right]\nonumber\\
& &+\frac{\hbar}{2m} (\partial_\lambda^x \tilde F_{\mu\nu}) \partial^\lambda_p \left[p^\nu \mathcal{F}\delta\left(p^2-m^2\right)\right]\,,
\end{eqnarray}
with the constraint condition,
\begin{eqnarray}
\label{A-condition}
p\cdot  \mathcal{A} \delta(p^2-m^2) &=& 0.
\end{eqnarray}
The integrated kinetic equations  and the constraint condition for the particle by integrating $p_0$ from
$0$ to $\infty$  are given by
\begin{eqnarray}
\label{Eq-F-a}
&&p\cdot \nabla  \mathcal{F}
=-\frac{\hbar\, p^\mu  }{2m} \left[ \frac{\tilde F_{\mu\nu}\bar  p^\lambda \nabla_\lambda }{E_p^2}
- (\bar \partial^\lambda_x \tilde F_{\mu\nu}) \bar \partial_\lambda^p \right]\mathcal{A}^{\nu}\,, \\
\label{Eq-A-a}
&&p\cdot  \nabla  \mathcal{A}_{\mu}
=F_{\mu\nu} \mathcal{A}^{\nu}
-\frac{\hbar\, p^\nu }{2m }\left[\frac{ \tilde F_{\mu\nu}  \bar p^\lambda  \nabla_\lambda}{E_p^2}
-(\bar\partial^\lambda_x \tilde F_{\mu\nu}) \bar\partial_\lambda^p \right]\mathcal{F}\,, \\
&&p \cdot  \mathcal{A} =0\,,
\end{eqnarray}
where $p$ has been now put on the mass-shell, i.e., $p_0=E_p$,
the derivative with respect to $p_0$ in $\nabla^\mu$ has been removed,
and $\bar p = p - (n\cdot p) n$ (when $n^\mu=(1,{\bf 0})$, $\bar{p}^\mu= (0,{\bf p})$).
To arrive at Eqs. (\ref{Eq-F-a},\ref{Eq-A-a}),
we have set $p_0=E_p$ before integration over $p_0$
so that $p_0$ derivatives are vanishing in $\mathcal{F}$ and $\mathcal{A}_{\mu}$.
Sometime it is useful to express equations in three-dimensions form
\begin{eqnarray}
\label{Eq-F-3v}
\left(\nabla_t +{\mathbf v}\cdot{\pmb{\nabla}}  \right)\mathcal{F}
&=&-\frac{\hbar }{2m E_p}\left[({\bf B}+{\bf E}\times {\bf v})
(  {\bf v}\cdot {\pmb \nabla} + E_p  \overleftarrow{\pmb \nabla}_x\cdot {\pmb \nabla}_p )\right.\nonumber\\
& &\left. -({\bf B}\cdot {\bf v})(  {\bf v}\cdot {\pmb \nabla} + E_p \overleftarrow{\pmb \nabla}_x\cdot {\pmb \nabla}_p){\bf v}\right]
\cdot\pmb{ \mathcal{A} }\,,\\
\label{Eq-A-3v}
\left(\nabla_t +{\mathbf v}\cdot{\pmb{\nabla}}  \right){\pmb{\mathcal{A}}}
&=& -{\bf E}({\bf v}\cdot {\pmb{\mathcal{A}}})+ {\bf B}\times\pmb{\mathcal{A}} \nonumber\\
& &- \frac{\hbar\, }{2m E_p }({\bf B}+{\bf E}\times {\bf v})
(  {\bf v}\cdot {\pmb \nabla} + E_p  \overleftarrow{\pmb \nabla}_x\cdot {\pmb \nabla}_p )
\mathcal{F}\,,
\end{eqnarray}
where ${\bf v}={\bf p}/E_p$, $\nabla_t=\partial_t + {\bf E}\cdot {\pmb{\nabla}_p}$,
${\pmb{\nabla}}=\pmb{\nabla}_x + {\bf B}\times {\pmb{\nabla}_p}$, and $\overleftarrow{\pmb \nabla}_x$
acts only on the electromagnetic fields on its left. Note that only in three-dimensions form, we are left with four independent functions $\mathcal{F}$ and $\pmb{\mathcal{A}}$ satisfying four kinetic equations.
The time component $\mathcal{A}^0={\bf v}\cdot\pmb{ \mathcal{A} }$ is not an independent function any more.
Equations (\ref{Eq-F-3v},\ref{Eq-A-3v}) reproduce the usual relativistic Vlasov equation and
Bargmann-Michel-Telegdi equation \cite{Bargmann:1959gz} for spin precession
in electromagnetic fields at classical limit.
In quantum correction terms $\mathcal{F}$ and ${\pmb{\mathcal{A}}}$ are coupled.
These terms contribute when $\mathcal{F}$ or ${\pmb{\mathcal{A}}}$  is inhomogeneous in phase space.
The terms with $\overleftarrow{\pmb \nabla}_x$ in Eq.(\ref{Eq-F-3v}) is associated with
the spin separation effect in Stern-Gerlach experiment.


\section{Chiral and spin effects}
\label{sec-chiral}
In relativistic non-central heavy-ion collisions, huge magnetic fields \cite{Bzdak:2011yy,Deng:2012pc,Bloczynski:2012en} and orbital angular momenta \cite{Liang:2004ph,Gao:2007bc,Becattini:2007sr} are generated with respect to the direction of the reaction plane. The initial orbital angular momentum can be further converted to vorticity fields in the fluid \cite{Csernai:2013bqa,Jiang:2016woz,Deng:2016gyh,Pang:2016igs}.
The huge magnetic and vorticity fields provide special probes to strong interaction matter in heavy-ion collisions.
They give rise to to many emerging phenomena, such as CME, CVE,CSE, GPE and LPE.
The CME, CVE and CSE originate from chiral anomaly, while GPE and LPE originate from spin-orbit couplings in particle scatterings. All these chiral and spin effects are of quantum nature, which can be well described by quantum kinetic theory based on Wigner functions. We give an overview about an application of Wigner functions to chiral and spin effects in this section.
{These results are scattered in different  Refs. \citen{Gao:2012ix,Gao:2015zka,Gao:2017gfq,Gao:2018jsi,Gao:2019znl,Gao:2019zhk},
we integrate them in this section. }
\subsection{Chiral anomaly}
\label{chiral-anomaly}
Chiral anomaly is a novel quantum effect which bridge ultraviolet and infrared properties of quantum fields \cite{Adler:1969gk,Bell:1969ts,Gribov:1981ku}.
Chiral anomaly can be naturally described in the Wigner function formalism. {Here we   derive the chiral anomaly for massless fermions and then for massive fermions
based on Ref.~\citen{Gao:2019zhk}.}

\subsubsection{Massless Fermions}
Using equations in Sect. \ref{subsec-massless}, we obtain the divergence of the axial current by integrating Eq. (\ref{ccke}) over four-momentum
\begin{equation}
\label{chiral-anomaly-massless}
\partial_\mu^x j_5^\mu = \sum_{s=\pm1} s \int d^4 p\, \partial^x_\mu{\mathscr{J}}^\mu_s
=F_{\mu\nu}  \sum_{s=\pm1} s \int d^4 p\, \partial^\nu_p {\mathscr{J}}^\mu_s \,.
\end{equation}
Since the last term is total derivative, it would vanish if ${\mathscr{J}}^\mu_s$ is a normal function of momentum, i.e., approaching to zero at infinite momentum. However as we mentioned before, the equations for Wigner functions are satisfied only for Wigner functions without normal ordering. In this case a singular term from vacuum or Dirac sea appears. This vacuum term gives a non-vanishing total derivative in momentum space and then chiral anomaly. To see this, it is convenient to rewrite ${\mathscr{J}}^\mu_s$ in Eq. (\ref{Jmu-1}) as
\begin{eqnarray}
{\mathscr{J}}^\mu_s &=& p^\mu f_s \delta\left(p^2\right)
+s \hbar\tilde F^{\mu\nu} p_\nu  f_s \delta'\left(p^2\right)
 -\frac{s\hbar}{2p_n} \epsilon^{\mu\nu\rho\sigma} n_\nu  p_\rho
\left(  \nabla_\sigma  f_s \right)\delta\left(p^2\right).
\end{eqnarray}
The vacuum contributions in $f_s$ from the first term does not contribute
because the contributions from left-hand and right-hand fermions cancel in Eq. (\ref{chiral-anomaly-massless}).
The vacuum contribution from the last term does not contribute either because of the derivative $\nabla_\sigma$.
Only the middle term contributes
\begin{equation}
\partial_\mu^x j_5^\mu
=-\frac{\hbar}{8\pi^{2}}F_{\mu\nu}\tilde{F}^{\mu\nu}C_{\mathrm{v}}\,,
\end{equation}
with $C_{\mathrm{v}}$ defined by
\begin{equation}
C_{\mathrm{v}}=\frac{1}{2\pi}\int d^{4}p\partial^{\mu}[p_{\mu}\delta^{\prime}(p^{2})]\,.
\end{equation}
We can evaluate the momentum integral directly by integrating over $p_0$ and obtain
\begin{equation}
C_{\mathrm{v}} = \int \frac{d^3{\bf p}}{2\pi}{\pmb \nabla}_{ p}\cdot {\pmb \Omega}_p =1\,,
\end{equation}
where ${\pmb \Omega}_p= {\bf p }/(2|{\bf p}|^3)$ is the Berry curvature in three-momentum.
To arrive at the final result, we have used the Gauss theorem in three-momentum or the identity
${\pmb \nabla}_{ p}\cdot {\pmb \Omega}_p=2\pi \delta^3({\bf p})$. Actually we can also finish
integration by using the regularization method
\begin{equation}
\delta'(x)=\frac{1}{\pi}\textrm{Im}\frac{1}{(x+i\epsilon)^2}\,,
\end{equation}
followed by Wick rotation and obtain
\begin{equation}
C_{\mathrm{v}}  =  \frac{1}{2\pi^2}\textrm{Im}\int d^{4}p\, \partial^{\mu}\left[\frac{p_{\mu}}{(p^2+i\epsilon)^2}\right]
=\frac{1}{2\pi^2}\int d^{4}p_E\, \partial_{\mu}\left(\frac{p^{\mu}_E}{p_E^4}\right) =1\,,
\end{equation}
where we have used the Gauss theorem in four-momentum or the identity
$\partial_{\mu}({p^{\mu}_E}/{p_E^4})=2\pi^2 \delta^4(p_E)$.
It is obvious that $p_{\mu}\delta^{\prime}(p^{2})$ plays the role of the Berry curvature of a four-dimensions monopole in Euclidean momentum space \cite{Chen:2012ca}. For massless fermions, we note that only the vacuum or Dirac sea contribution gives rise to chiral anomaly in the form of four-dimensions or three-dimensions Berry curvatures.

\subsubsection{Massive Fermions}
For massive fermions, we use the result in Sect. \ref{subsec-massive}.  The divergence of the axial current can be obtained by integrating Eq. (\ref{P}) over $p$ after substituting Eq. (\ref{A-1-b}) into it,
\begin{equation}
\label{chiral_anomaly}
\partial^\mu_x j_\mu^{5}= - \frac{2 m}{\hbar} j_5  - \frac{\hbar}{8\pi^2} C_\textrm{v} F^{\mu\nu} \tilde F_{\mu\nu}\,,
\end{equation}
where
\begin{eqnarray}
j_5 = \int d^4 p \mathscr{P},\ \ \ \ \
C_{\mathrm{v}}  =
\frac{1}{2\pi}\int d^{4}p\partial^{\mu}[p_{\mu}\delta^{\prime}(p^{2}-m^2)]\,.
\label{eq:h12}
\end{eqnarray}
Again we can evaluate this integral directly
\begin{equation}
C_{\mathrm{v}} = \int \frac{d^3{\bf p}}{2\pi}{\mathbf \partial}_{\bf p}\cdot
\left[ \, \frac{\hat{\bf p}}{2({\bf p}^2+m^2)}\right] =1 \,,
\end{equation}
or evaluate the integral by Wick rotation
\begin{eqnarray}
C_{\mathrm{v}}
&=&\frac{1}{2\pi^2}\textrm{Im}\int d^{4}p\, \partial^{\mu}\left[\frac{p_{\mu}}{(p^2-m^2+i\epsilon)^2}\right] \nonumber\\
&=&\frac{1}{2\pi^2}\int d^{4}p_E\, \partial_{\mu}\left[\frac{p^{\mu}_E}{(p_E^2+m^2)^2}\right] =1\,,
\end{eqnarray}
where we have used Gauss theorems in momentum space of three-dimensions and four-dimensions, respectively.
Here we can define the Berry curvature for massive fermions as $\hat{\bf p}/(2{E_p}^2)$
with $E_p=\sqrt{{\bf p}^2 + m^2}$ in three-momentum. Note that there is no genuine singularity at ${\bf p}=0$
in the Berry curvature, it behaves as a Berry monopole approximately at large momentum
when the fermion's mass is negligible.

Although chiral anomaly is related to the Berry curvature at chiral limit,
it is different from the result of Refs.~\citen{Stephanov:2012ki,Manuel:2014dza,Chen:2012ca}
in which chiral anomaly is proportional to the distribution function at zero momentum.
In our present approach, chiral anomaly for either massless or massive fermions is universal and
is independent of normal phase space distributions at zero momentum.

Now let us verify conservation of the vector current. As an example,
we consider massive fermions since the massless case can be similarly obtained.
Taking the space-time divergence of both sides of Eq. (\ref{Vmu}) and an integration over $p$, we obtain
\begin{eqnarray}
\label{V-conservation}
\partial^\mu_x j_\mu &=& \int d^4 p \partial^\mu_x \mathscr{V}_{\mu} \nonumber \\
&=&\int d^4 p\left\{ \frac{p_\mu}{m} \partial^\mu_x \mathscr{F}
+\frac{\hbar}{2m^2}\epsilon_{\mu\nu\rho\sigma}\partial^\mu_x \left[F^{\nu\lambda}\partial_\lambda^p
\left( p^\rho \mathscr{A}^{\sigma}\right)\right]\right\}\,.
\end{eqnarray}
Since the vacuum contribution for different spin states are the same,
there is no net vacuum contribution from $\mathscr{A}^{\sigma}$ due to cancelation of different spin states.
The absence of vacuum contribution makes the total momentum derivative in Eq. (\ref{V-conservation}) vanish.
After dropping this term and using Eqs. (\ref{PS-1-a1},\ref{PS-2-a1}), we have
\begin{equation}
\partial^\mu_x j_\mu =  \int d^4 p \left[\frac{1}{m} F_{\mu\nu}\partial^\nu_p ( p^\mu \mathscr{F})
 +\frac{\hbar}{2m^2} \Delta \tilde F_{\mu\nu}( p^\mu \mathscr{A}^\nu)\right]\,,
\end{equation}
where we have used the Maxwell's equation $\partial^\mu_x \tilde F_{\mu\nu}=0$ during moving $p^\mu$ through the operator $\Delta$. Again because there is no vacuum contribution from $\mathscr{A}^{\sigma}$,
only the first term has vacuum contribution which gives
\begin{eqnarray}
\partial^\mu_x j_\mu &=&- \frac{1}{2\pi^3}  F_{\mu\nu} \int d^4 p\,
\partial^\nu_p [ p^\mu \delta(p^2-m^2)\theta(-p_0)]=0 \,.
\end{eqnarray}
We see that the Dirac sea or vacuum contribution does not influence the conservation law
for the electric charge as expected.


\subsection{Currents induced by magnetic field}
As we all know that the electromagnetic field can induce electric currents in classical physics, e.g., the Olm's current and Hall's current. Electromagnetic fields can induce currents from quantum effects, such as the CME or CSE. In this subsection we give a brief overview on derivation of these effects in the Wigner function approach {based on the works \citen{Gao:2012ix,Gao:2017gfq,Gao:2018jsi,Gao:2019znl}}. First we restrict ourselves to a system of massless fermions in uniform electromagnetic field with $\partial_x^\lambda F^{\mu\nu}=0$ and then we discuss possible mass corrections to the results for massless fermions.

We consider a system of massless fermions near equilibrium and choose $f^{(0)}$ in Eq. (\ref{Jn-0})
as the Fermi-Dirac distribution in global equilibrium
\begin{eqnarray}
\label{f0s}
f^{(0)}_s &=& \frac{1}{4\pi^3} \left[
\frac{1}{e^{\beta\cdot p-\bar\mu_s }+1}\frac{\delta(p_0-|{\bf p}|)}{2 |{\bf p}|}\right.\nonumber\\
&& \hspace{1cm}\left. + \left( \frac{1}{e^{-\beta\cdot p + \bar\mu_s }+1}-1\right) \frac{\delta(p_0+|{\bf p}|)}{2 |{\bf p}|}\right]\,.
\end{eqnarray}
or in a covariant form
\begin{eqnarray}
\label{f0s-covariant}
f^{(0)}_s &=& \frac{1}{4\pi^3 } \left[
 \theta(p_0)\frac{1}{e^{\beta\cdot p-\bar\mu_s }+1}
 +\theta(-p_0)\left( \frac{1}{e^{-\beta\cdot p + \bar\mu_s }+1}-1\right) \right]\delta(p^2)\;.
\end{eqnarray}
where $\beta^\mu = {u^\mu}/{T}$ and $\bar \mu_s = {\mu_s}/{T} = \bar\mu + s \bar \mu_5$
with $T$ being the temperature, $u^\mu$ being the four-velocity of the fluid, and $\mu_s$ ($s=\pm$), $\mu$ and $\mu_5$ being the right-hand/left-hand, vector and chiral chemical potentials, respectively.
The distribution function (\ref{f0s-covariant}) must satisfy Eq. (\ref{kinetic-eq-0}).
One can verify that the distribution function (\ref{f0s-covariant})
indeed satisfies Eq. (\ref{kinetic-eq-0}) under conditions
\begin{eqnarray}
\label{pd1}
\partial_\mu \beta_\nu +\partial_\nu\beta_\mu &=& 0\,,\\
\label{pd3}
\partial_\mu \bar\mu + F_{\mu\nu} \beta^\nu &=& 0 \,,\\
\label{pd2}
\partial_\mu\bar\mu_5 &=& 0 \,.
\end{eqnarray}
Since we are only concerned with electromagnetic effects,
we assume that $u^\mu$ is a constant vector. Up to first order in $\hbar$ the chiral Wigner function reads
\begin{eqnarray}
{\mathscr{J}}^{(1)\mu}_s &=& p^\mu f_s^{(1)} \delta(p^2)
+s \tilde F^{\mu\nu} p_\nu  f_s^{(0)} \delta'(p^2)\nonumber\\
&&-\frac{s}{2p_n} \epsilon^{\mu\nu\rho\sigma} n_\nu  p_\rho  (  \nabla_\sigma  f_s^{(0)})\delta(p^2)\,,
\label{first-order-cwf}
\end{eqnarray}
The last term vanishes under conditions (\ref{pd1}-\ref{pd2}). Since the second term does not depend on the auxiliary vector $n^\mu$, the first term must be independent of $n^\mu$ either.
In order to see the response from electromagnetic fields we can simply set $f_s^{(1)}=0$ here.
So Eq. (\ref{first-order-cwf}) becomes
\begin{equation}
\label{CVE-J}
{\mathscr{J}}^{(1)\mu}_s = s \tilde F^{\mu\nu} p_\nu  f_s^{(0)} \delta'(p^2) \,.
\end{equation}
Integrating the above over $p$ gives rise to the first order current
\begin{equation}
\label{js-1-a}
j^{(1)\mu}_s = \xi_{Bs} B^\mu \;,
\end{equation}
with the coefficient
\begin{equation}
\label{xibs}
\xi_{Bs} = \frac{s}{4\pi^2}\mu_s \;.
\end{equation}
Note that the magnetic field in the local frame is defined as $B^\mu=\tilde F^{\mu\nu} u_\nu$ through the fluid velocity $u_\nu$ instead of $n_\nu$. It follows that the vector and axial current are given by
\begin{eqnarray}
\label{j-1-a}
j^{(1)\mu} &=&  \xi_B B^\mu \,,\\
\label{j5-1-a}
j^{(1)\mu}_5 &=& \xi_{B5} B^\mu \,,
\end{eqnarray}
with anomalous transport coefficients $\xi_B$ and $\xi_{B5}$ being associated with CME and CSE respectively and given by
\begin{eqnarray}
\xi_B = \frac{\mu_5}{2\pi^2}, \,\,\,\,  \xi_{B5} = \frac{\mu}{2\pi^2} \,.
\label{n-n5-xi-xi5}
\end{eqnarray}
We see that the CME current is proportional to the chiral chemical potential $\mu_5$
while the CSE one is proportional to the vector chemical potential $\mu$.

When the fermion has a finite mass, there is no problem for introducing the vector chemical potential $\mu$, but there will be issues for the chiral chemical potential $\mu_5$. Hence we consider the CSE for massive fermions in which only $\mu$ is relevant.{\cite{Gao:2019znl}} We consider a global equilibrium solution of $\mathscr{A}_{\mu}$ and $\mathscr{F}$ in Eqs. (\ref{F-0-b},\ref{A-0-b},\ref{A-1-b}) with
\begin{eqnarray}
\label{A-distribution}
&&\mathcal{A}_\mu^{(0)}=\mathcal{A}_\mu^{(1)}=0,\\
\label{FD-distribution}
&& \mathcal{F}^{(0)}=\frac{m}{2\pi^3}
\left[\frac{\theta\left(u\cdot p\right)}{e^{{( u\cdot p-\mu)}/{T}}+1}
+\frac{\theta\left(-u\cdot p\right)}{e^{-{(u\cdot  p-\mu)}/{T}}+1}\right] \,,
\end{eqnarray}
Substituting them into Eq. (\ref{A-1-b}) and integrating over $p$, we obtain the chiral current
\begin{equation}
\label{CSE}
j_5^{(1)\mu }=\int d^4 p \mathscr{A}^{(1)\mu} =  \xi_{B5} B^\mu \,,
\end{equation}
where the CSE coefficient is
\begin{equation}
\xi_{B5}= \frac{1}{2\pi^2} \int_0^\infty d p
 \left(\frac{1}{e^{(E_p-\mu)/T}+1}- \frac{1}{e^{(E_p+\mu)/T}+1}\right) \,.
\end{equation}
which is consistent with the result from other methods~\cite{Lin:2018aon}.
At zero temperature, we have the analytic result
\begin{eqnarray}
\xi_{B5}|_{T\to 0}=\frac{\mu}{2\pi^2}\sqrt{1-\frac{m^2}{\mu^2}}\,.
\end{eqnarray}

For massive fermions, we can also obtain the magnetic moment density {\cite{Gao:2019znl}}
which is determined by the spatial components of $\mathscr{S}_{\mu\nu}$ in the rest frame of $u^\mu$.
We insert the result of $ \mathscr{A}_{\mu}^{(1)}$ in Eq. (\ref{A-1-b}) into Eq. (\ref{S-1-a}) to get $\mathscr{S}_{\mu\nu}^{(1)}$.
It follows that the magnetic moment density or magnetization vector is given by
\begin{equation}
\label{MVE}
M_\mu^{(1)} =\frac{1}{2}\epsilon_{\nu\mu\alpha\beta} u^\nu \int d^4 p \mathscr{S}^{(1)\alpha\beta}
= \kappa B_\mu \,,
\end{equation}
where the magnetic susceptibility $\kappa$ is given by
\begin{equation}
\kappa=\frac{m}{2\pi^2}\int_0^\infty\frac{d p}{E_p} \left(\frac{1}{e^{(E_p-\mu)/T}+1}
+ \frac{1}{e^{(E_p+\mu)/T}+1}\right)\,.
\label{m-susceptibility}
\end{equation}
At zero temperature, we can obtain an analytic expression
\begin{eqnarray}
\kappa|_{T\to 0}&=&\frac{m}{2\pi^2}\ln \frac{|\mu|+\sqrt{\mu^2-m^2}}{m}\,.
\end{eqnarray}
We see in (\ref{m-susceptibility}) that the magnetic susceptibility and then magnetic moment density
are vanishing at massless limit.


\subsection{Chiral magnetic conductivity for non-zero frequencies}
The results we presented in the preceding section is only valid at the static limit with zero frequency. In the Wigner function formalism, we can also derive a general chiral magnetic conductivity at non-zero frequencies.{\cite{Gao:2015zka}}
In order do this, we expand equations in electromagnetic fields instead of $\hbar$
and keep full space-time derivatives on fields.
We can still write the chiral Wigner functions as a sum of the zeroth-order and first-order contribution
in the expansion of electromagnetic fields,
\begin{eqnarray}
\mathscr{J}_s^\mu &=&\mathscr{J}^{(0)\mu}_s+\mathscr{J}^{(1)\mu}_s\,.
\end{eqnarray}
Here the first-order contribution $\mathscr{J}^{(1)\mu}_s$ includes all possible derivative terms
(with all possible powers of $\hbar$). At the zeroth order the set of equations
for chiral Wigner functions are given by
\begin{eqnarray}
\label{Js-c1-F-0}
p^\mu\mathscr{J}^{(0)}_{s\mu} &=&0\,, \\
\label{Js-eq-F-0}
\partial_x^\mu\mathscr{J}^{(0)}_{s\mu} &=& 0\,, \\
\label{Js-c2-F-0}
\hbar\epsilon_{\mu\nu\rho\sigma}\partial_x^\rho \mathscr{J}^{(0)\sigma}_s
&=&-2s\left(p_\mu \mathscr{J}^{(0)}_{s\nu}-p_\nu \mathscr{J}^{(0)}_{s\mu}\right)\,.
\end{eqnarray}
We assume the zeroth order solution takes the form
\begin{equation}
\label{J-0}
\mathscr{J}^{(0)\mu}_s = p^\mu f^{(0)}_s\delta(p^2)
\end{equation}
where $f^{(0)}_s$ is the Fermi-Dirac distribution given in Eqs. (\ref{f0s},\ref{f0s-covariant}).
Different from conditions in Eqs. (\ref{pd1}-\ref{pd2}), we assume thermal quantities
$u^\mu$, $\mu$ and $\mu_5$ are all constants.
The first order equations read
\begin{eqnarray}
\label{Js-c1-F-1}
& &p^\mu\mathscr{J}_{s\mu}^{(1)}  -\frac{1}{2}\hbar  j_1\left(\frac{1}{2}\hbar\Delta\right)F^{\mu\nu}\partial_\nu^p \mathscr{J}_{s\mu}^{(0)} =0,\\
\label{Js-eq-F-1}
& &\partial_x^\mu\mathscr{J}_{s\mu}^{(1)}  - j_0\left(\frac{1}{2}\hbar\Delta\right)F^{\mu\nu}\partial_\nu^p \mathscr{J}_{s\mu}^{(0)} = 0,\\
\label{Js-c2-F-1}
& &\hbar\epsilon_{\mu\nu\rho\sigma}\left[\partial_x^\rho \mathscr{J}^{(1)\sigma}_s
- j_0\left(\frac{1}{2}\hbar\Delta\right)F^{\rho\lambda}\partial_\lambda^p\mathscr{J}^{(0)\sigma}_s  \right]\nonumber\\
&&=-2s\left(p_\mu \mathscr{J}^{(1)}_{s\nu}-p_\nu \mathscr{J}^{(1)}_{s\mu}\right)\nonumber \\
&&+s\hbar j_1\left(\frac{1}{2}\hbar\Delta\right)\left[ F_{\mu\lambda}\partial^\lambda_p \mathscr{J}_{s\nu}^{(0)}
- F_{\nu\lambda}\partial^\lambda_p \mathscr{J}_{s\mu}^{(0)}  \right]\,.
\end{eqnarray}
Contracting $\partial^\nu_x $ with Eq. (\ref{Js-c2-F-1}) and using the Eq. (\ref{Js-eq-F-1}), we have
\begin{eqnarray}
p_\nu \partial^\nu_x  \mathscr{J}^{(1)}_{s\mu} &=&-\frac{s}{2}\hbar
\epsilon_{\mu\nu\rho\sigma} \partial^\nu_x \left[ j_0\left(\frac{1}{2}\hbar\Delta\right)F^{\rho\lambda}\partial_\lambda^p\mathscr{J}^{(0)\sigma}_s\right]\nonumber\\
&&+p_\mu j_0\left(\frac{1}{2}\hbar\Delta\right)F^{\nu\lambda}\partial_\lambda^p \mathscr{J}_{s\nu}^{(0)}\nonumber\\
& &-\frac{1}{2}\hbar \partial^\nu_x \left[ j_1\left(\frac{1}{2}\hbar\Delta\right)\left( F_{\mu\lambda}\partial^\lambda_p \mathscr{J}_{s\nu}^{(0)}
- F_{\nu\lambda}\partial^\lambda_p \mathscr{J}_{s\mu}^{(0)}  \right)\right] \,.
\end{eqnarray}
A formal solution to the above equation is given by
\begin{eqnarray}
\mathscr{J}^{(1)}_{s\mu }&=&X_\mu - \frac{s}{2\hat Q_1}\hbar
\epsilon_{\mu\nu\rho\sigma} \partial^\nu_x \left[ j_0\left(\frac{1}{2}\hbar\Delta\right)F^{\rho\lambda}\partial_\lambda^p\mathscr{J}^{(0)\sigma}_s\right]\nonumber\\
&&+ \frac{1}{\hat Q_1 }p_\mu j_0\left(\frac{1}{2}\hbar\Delta\right)F^{\nu\lambda}\partial_\lambda^p \mathscr{J}_{s\nu}^{(0)}\nonumber\\
&&-\frac{1}{2\hat Q_1}\hbar \partial^\nu_x \left[ j_1\left(\frac{1}{2}\hbar\Delta\right)\left( F_{\mu\lambda}\partial^\lambda_p \mathscr{J}_{s\nu}^{(0)}
- F_{\nu\lambda}\partial^\lambda_p \mathscr{J}_{s\mu}^{(0)}  \right)\right]\,,
\label{solution-em}
\end{eqnarray}
where $\hat Q_1\equiv p\cdot \partial_x $ and $X_\mu$ is an arbitrary vector satisfying
$\hat Q_1 X_\mu =0$ which is assumed to be vanishing in the following.
We can demonstrate that this result (\ref{solution-em}) satisfies Eqs. (\ref{Js-c1-F-1}-\ref{Js-c2-F-1}).
It is more convenient to rewrite it in momentum space by replacing
$\hbar\partial_{x}\rightarrow-ik$ and $\Delta\rightarrow-ik\cdot\partial_{p}$,
\begin{eqnarray}
\mathscr{J}_{s\mu}^{(1)}(k,p) & = &  -i\frac{s{\hbar}}{2p\cdot k}\epsilon_{\mu\nu\rho\sigma}k^{\nu}p^{\sigma}A^{\rho}j_{0}
\left(-\frac{i k\cdot \partial_p}{2}\right)(k\cdot\partial_{p})[f_{s}\delta(p^{2})]\nonumber \\
&& +\frac{{\hbar}}{p\cdot k}p_{\mu}[(p\cdot k)(A\cdot\partial_{p})-(p\cdot A)(k\cdot\partial_{p})]j_{0}\left(-\frac{i k\cdot \partial_p}{2}\right)[f_{s}\delta(p^{2})]\nonumber \\
&& +\frac{1}{4p\cdot k}[k_{\mu}(k\cdot A)-k^{2}A_{\mu}](k\cdot\partial_{p})j_{0}
\left(-\frac{i k\cdot \partial_p}{2}\right)[f_{s}\delta(p^{2})]\nonumber \\
&& +i\frac{{\hbar}}{2}[k_{\mu}(A\cdot\partial_{p})-A_{\mu}(k\cdot\partial_{p})]j_{1}
\left(-\frac{i k\cdot \partial_p}{2}\right)[f_{s}\delta(p^{2})]\,.
\end{eqnarray}
For the parity-odd part, we consider the first term, which can be rewritten as
\begin{eqnarray}
\mathscr{J}_{s\mu}^{(1)}(k,p)
& = & i\frac{s}{2p\cdot k}\epsilon_{\mu\nu\rho\sigma}k^{\nu}p^{\rho}A^{\sigma}\left\{ f_{s}\left(p+\frac{1}{2}k\right)\delta\left[\left(p+\frac{1}{2}k\right)^{2}\right]\right.\nonumber \\
&  & \left.-f_{s}\left(p-\frac{1}{2}k\right)\delta\left[\left(p-\frac{1}{2}k\right)^{2}\right]\right\}\,,
\label{eq:chiral-part}
\end{eqnarray}
where we have used the translation operator
\begin{equation}
\exp\left(\frac{1}{2}k\cdot\partial_{p}\right)f_{s}\delta(p^{2})
=f_{s}\left(p+\frac{1}{2}k\right)\delta\left[\left(p+\frac{1}{2}k\right)^{2}\right]\,.
\end{equation}
The current can be obtained by an integration over $p$
\begin{equation}
j_{s\mu}^{(1)}= \int d^4 p\mathscr{ J}_{s\mu}^{(1)}= i{\hbar}\epsilon_{\mu\nu\rho\sigma}u^\nu k^\rho A^\sigma \xi_{Bs}\,,
\end{equation}
where the chiral conductivity is given by
\begin{eqnarray}
\xi_{Bs} & = & -\frac{s k^2}{2 \bar k^2 }\int d^4p \frac{ u\cdot p}{p\cdot k}
\left\{f_s\left(p+\frac{k}{2}\right)\delta\left[\left(p+\frac{k}{2}\right)^2\right]\right. \nonumber\\
&& \left. -f_s\left(p-\frac{k}{2}\right)\delta\left[\left(p-\frac{k}{2}\right)^2\right]\right\} \,.
\end{eqnarray}
We set $u^\mu=(1,0,0,0)$ and carry out the integral
\begin{eqnarray}
\xi_{Bs} &=&\frac{s}{16\pi^2}\frac{ \left({{\bf k}^2}-\omega^2\right)}{ {|{\bf k}|^3}}\int{d{|{\bf p}|}}
\left\{\left(2 \left|\bf p \right|-{\omega}\right)
\ln\left[\frac{(\omega + i\epsilon -|{\bf p}|)^2 -(|{\bf p}|+|{\bf k}|)^2 }
{(\omega + i\epsilon -|{\bf p}|)^2 -(|{\bf p}|-|{\bf k}|)^2}\right]\right.\nonumber\\
&&\left. -\left( 2 \left|\bf p \right|+{\omega}\right)
\ln\left[\frac{(\omega + i\epsilon +|{\bf p}|)^2 -(|{\bf p}|-|{\bf k}|)^2 }
{(\omega + i\epsilon +|{\bf p}|)^2 -(|{\bf p}|+|{\bf k}|)^2}\right]\right\}\nonumber\\
&&\times\left[ \frac{1}{e^{\beta\left|\bf p\right|
-\bar\mu_s}+1}-\frac{1}{e^{\beta\left|\bf p\right| +\bar\mu_s}+1}\right]\,,
\end{eqnarray}
where we have introduced the $i\epsilon$ prescription.
The above result is just the one-loop result from quantum field theory \cite{Kharzeev:2009pj}.
It follows that the real part is given by
\begin{eqnarray}
\textrm{Re}\, \xi_{Bs}&=&\frac{s }{16\pi^2}\frac{ \left({{\bf k}^2}-\omega^2\right)}{ {|{\bf k}|^3}}\int{d{|{\bf p}|}}
\left\{\left(2 \left|\bf p \right|-{\omega}\right)
\ln\left|\frac{(\omega  -|{\bf p}|)^2 -(|{\bf p}|+|{\bf k}|)^2 }
{(\omega  -|{\bf p}|)^2 -(|{\bf p}|-|{\bf k}|)^2}\right|\right.\nonumber\\
& &\left.-\left( 2 \left|\bf p \right|+{\omega}\right)
\ln\left|\frac{(\omega  +|{\bf p}|)^2 -(|{\bf p}|-|{\bf k}|)^2 }
{(\omega  +|{\bf p}|)^2 -(|{\bf p}|+|{\bf k}|)^2}\right|\right\} \nonumber\\
&&\times \left[ \frac{1}{e^{\beta\left|\bf p\right| -\bar\mu_s}+1}-\frac{1}{e^{\beta\left|\bf p\right| +\bar\mu_s}+1}\right] \,,
\end{eqnarray}
and the imaginary part is given by
\begin{eqnarray}
\textrm{Im}\, \xi_{Bs}&=&\frac{s }{16\pi}\frac{ \left({{\bf k}^2}-\omega^2\right)}{ {|{\bf k}|^3}}\int{d{|{\bf p}|}}
\left\{\left(2 \left|\bf p \right|-{\omega}\right)
\left[\theta\left( {\bf k}^2-\omega^2\right)\theta\left(2|{\bf p}|-|{\bf k}|-\omega\right)\frac{}{}\right.\right.\nonumber\\
& & \left.\frac{}{} +\theta\left(\omega^2- {\bf k}^2\right)\theta\left(2|{\bf p}|+|{\bf k}|-\omega\right)\theta\left(|{\bf k}|+\omega-2|{\bf p}|\right)\right]\nonumber\\
&&\times\left[\theta\left(\omega- |{\bf p}|\right)-\theta\left(|{\bf p}|-\omega\right)\right]\nonumber\\
& &\left.-\left( 2 \left|\bf p \right|+{\omega}\right)\left[\theta\left( {\bf k}^2-\omega^2\right)\theta\left(2|{\bf p}|-|{\bf k}| +\omega\right)\frac{}{}\right.\right.\nonumber\\
& &\left. \left.\frac{}{} +\theta\left(\omega^2- {\bf k}^2\right)\theta\left(|{\bf k}|-\omega -2|{\bf p}|\right)\theta\left( 2|{\bf p}|+|{\bf k}| + \omega \right)\right]\right\}\nonumber\\
&&\times \left[\theta\left(-\omega- |{\bf p}|\right)-\theta\left(|{\bf p}|+\omega\right)\right]
\left[ \frac{1}{e^{\beta\left|\bf p\right| -\bar\mu_s}+1}-\frac{1}{e^{\beta\left|\bf p\right| +\bar\mu_s}+1}\right]\,.
\end{eqnarray}
If we assume that external frequency and momentum are much smaller
than internal momentum, $\omega,|\mathbf{k}|\ll|\mathbf{p}|$, we
can reproduce the hard thermal loop (HTL) or hard dense loop (HDL) result
\cite{Son:2012zy,Manuel:2013zaa},
\begin{eqnarray}
\xi_{Bs}^{\mathrm{HTL/HDL}}(\omega,\mathbf{k}) & = &
 \frac{s }{4\pi^2 }\mu_s
 \left(1-\frac{\omega^{2}}{|\mathbf{k}|^{2}}\right)
 \left[1-\frac{\omega}{2|\mathbf{k}|}\ln\frac{\omega+|\mathbf{k}|}{\omega-|\mathbf{k}|}\right]\,,
\end{eqnarray}
Then in the limit $|{\bf k}|\rightarrow 0$, we can reproduce the result (\ref{xibs}) in the static limit.
We have neglected all the possible branch cuts in the above derivation.


\subsection{Currents induced by vorticity}
Similar to the magnetic field, in this subsection we give a derivation of vector and chiral currents generated by vorticity
in the Wigner function approach.{\cite{Gao:2017gfq,Gao:2018jsi}}
 For simplicity, we neglect the electromagnetic field. The zeroth order distribution function is still the Fermi-Dirac one, but the fluid velocity is not a constant.
With the condition (\ref{pd1}), the chiral system can have a global vorticity as
\begin{equation}
\beta_\mu =-\Omega_{\mu\nu}x^\nu,\,\,\,\,\
\Omega_{\mu\nu}=\frac{1}{2}\left(\partial_\mu \beta_\nu -\partial_\nu \beta_\mu\right)\;,
\end{equation}
where $\Omega^{\mu\nu}$ is a constant antisymmetric tensor. The first-order solution has the form
\begin{equation}
\mathscr{J}_{s}^{(1)\mu} = p^{\mu}f^{(1)}_s\delta(p^{2})
-\frac{s }{2n\cdot p}\epsilon^{\mu\nu\rho\sigma} n _\nu
p_\rho (\partial^x_\sigma f^{(0)}_s) \delta\left(p^2\right)\,,
\end{equation}
where the second term depends on $n^{\mu}$ implying that the first term must also depend on $n^{\mu}$
because $\mathscr{J}_{s}^{(1)\mu}$ should not depend on it.
Hence we cannot simply set $f^{(1)}_s=0$ {like we did in deriving the CME in Eq.(\ref{CVE-J})}. We need to extract the $n^\mu$ dependent part with
the help of the transformation rule derived in the previous section.
Substituting the Fermi-Dirac distribution into the transformation (\ref{df1}) leads to
\begin{eqnarray}
\delta f^{(1)}_s & = & -s\frac{p^{\mu}\epsilon^{\lambda\nu\rho\sigma}n_{\lambda}n_{\nu}^{\prime}p_{\sigma}\partial_{\rho}^{x}\beta_{\mu}}{2\left(n^{\prime}\cdot p\right)\left(n\cdot p\right)}\frac{df^{(0)}_s}{d(\beta\cdot p)}\nonumber \\
 & = & -s\frac{n_{\alpha}^{\prime}p_{\gamma}\tilde{\Omega}^{\alpha\gamma}}{2\left(n^{\prime}\cdot p\right)}\frac{df^{(0)}_s}{d(\beta\cdot p)}+s\frac{n_{\alpha}p_{\gamma}\tilde{\Omega}^{\alpha\gamma}}{2\left(n\cdot p\right)}\frac{df^{(0)}_s}{d(\beta\cdot p)}\,,
\end{eqnarray}
where we have used the dual vorticity tensor $\tilde \Omega^{\mu\nu}=\epsilon^{\mu\nu\rho\sigma}\Omega_{\rho\sigma}/2$.
We can express $f^{(1)}_s$ as
\begin{equation}
f^{(1)}_s = \tilde{f}^{(1)}_s - s\frac{n_{\alpha}p_{\gamma}\tilde{\Omega}^{\alpha\gamma}}{2\left(n\cdot p\right)}\frac{df^{(0)}_s}{d(\beta\cdot p)}\,,
\label{f1}
\end{equation}
where $\tilde{f}^{(1)}_s$ is independent of $n^{\mu}$, i.e. $\delta\tilde{f}^{(1)}_s=0$.
We can choose a specific solution with $\tilde{f}^{(1)}_s=0$.
Substituting Eq. (\ref{f1}) into Eq. (\ref{Jmu-1}) yields
\begin{eqnarray}
\mathscr{J}^{(1)\mu}_s & = & -p^{\mu}\frac{s}{2 n\cdot p}n_{\alpha}p_{\gamma}\tilde{\Omega}^{\alpha\gamma}
\frac{df^{(0)}_s}{d(\beta\cdot p)}\delta(p^{2})\nonumber\\
&&-\frac{s}{2n\cdot p}p^\lambda \epsilon^{\mu\nu\rho\sigma}n_{\nu}p_{\rho} \Omega_{\sigma\lambda }\frac{df^{(0)}_s}{d(\beta\cdot p)} \delta(p^{2})\nonumber \\
& = & -\frac{s}{2}\tilde{\Omega}^{\mu\nu}p_{\nu}\frac{df^{(0)}_s}{d(\beta\cdot p)}\delta(p^{2})\,,
\label{eq:j-mu-1-inv}
\end{eqnarray}
where we see in the last equality that $\mathscr{J}^{(1)\mu}_s$ is independent of $n^{\mu}$.
Integrating (\ref{eq:j-mu-1-inv}) over $p$ leads to
\begin{equation}
\label{js-1-a}
j^{(1)\mu}_s = \int d^4p \mathscr{J}^{(1)\mu}_s = \xi_s \omega^\mu\,,
\end{equation}
where
\begin{eqnarray}
\label{xis}
\xi_s = \frac{s }{12\pi^2}\left(\pi^2T^2+3\mu_s^2\frac{}{}\right),
\ \ \ \ \ \omega^\mu =\tilde \Omega^{\mu\nu}u_\nu \,.
\end{eqnarray}
It follows that the vector and chiral current are given by
\begin{eqnarray}
\label{j-1-a}
j^{(1)\mu} &=& \xi \omega^\mu \equiv \frac{\mu\mu_5 }{\pi^2}\omega^\mu \;,\\
\label{j5-1-a}
j^{(1)\mu}_5 &=&  \xi_5 \omega^\mu \equiv
\frac{1}{6\pi^2}\left[\pi^2T^2+3(\mu^2 + \mu_5^2)\frac{}{}\right] \omega^\mu \;,
\end{eqnarray}
where $\xi$ and $\xi_5$ are the anomalous transport coefficients for CVE and LPE, respectively.

We can verify that  Eq. (\ref{eq:j-mu-1-inv}) is still valid when $n^\mu$ depends on space-time.
If we choose $n^\mu \equiv u^\mu$, the two terms in the right-hand side of
the first equality in Eq. (\ref{eq:j-mu-1-inv}),
which we call $j_{s}^{(1)\mu}(1)$ and $j_{s}^{(1)\mu}(2)$, are evaluated as
\begin{eqnarray}
j_{s}^{(1)\mu}(1) & = & -\frac{s}{2}u_{\alpha}\tilde{\Omega}^{\alpha\gamma}
\int d^{4}p\frac{1}{u\cdot p}p^{\mu}p_{\gamma}\frac{df_{s}}{d(\beta\cdot p)}\delta(p^{2})
=\frac{1}{3}T\xi_{s}\omega^{\mu}\,,\nonumber \\
j_{s}^{(1)\mu}(2) & = & -\frac{s}{2}\epsilon^{\mu\nu\rho\sigma}u_{\nu}
\int d^{4}p\frac{1}{u\cdot p}p_{\rho}(\partial_{\sigma}^{x}f_{s})\delta(p^{2})
=\frac{2}{3}T\xi_{s}\omega^{\mu}\,.
\label{eq:j1-j2}
\end{eqnarray}
We see in Eq. (\ref{eq:j1-j2}) that $j_{s}^{(1)\mu}(1)$ and $j_{s}^{(1)\mu}(2)$ contribute
to the full CVE current by 1/3 and 2/3 repectively. In order to see
the physical meaning of $j_{s}^{(1)\mu}(1)$ and $j_{s}^{(1)\mu}(2)$,
we choose a local static frame $n^{\mu}=u^{\mu}=(1,0,0,0)$ at a specific
space-time point but with $\partial_{\mu}u_{\nu}\neq 0$ in its vicinity,
in which we can obtain the explicit form of $j_{s}^{\mu}(1)$ and
$j_{s}^{\mu}(2)$ in three spatial dimensions (3D),
\begin{eqnarray}
\mathbf{j}_{s}(1)&=&\mathbf{j}_{s}^{(0)}(1)+\hbar\mathbf{j}_{s}^{(1)}(1)
\approx  \int\frac{d^{3}{\bf p}}{(2\pi)^{3}}  \frac{\mathbf{p}}{|\mathbf{p}|} \nonumber\\
& &\times \left[f_{\mathrm{FD}}\left(\beta|\mathbf{p}|-\beta\mu_{s}-s\hbar\frac{\mathbf{p}\cdot\boldsymbol{\omega}}{2|\mathbf{p}|}\right) +f_{\mathrm{FD}}\left(\beta|\mathbf{p}|+\beta\mu_{s}-s\hbar\frac{\mathbf{p}\cdot\boldsymbol{\omega}}{2|\mathbf{p}|}\right)\right],\nonumber \\
\mathbf{j}_{s}(2) & = &\hbar \mathbf{j}_{s}^{(1)}(2)=\hbar\lim_{|\mathbf{v}|=0}\nabla\times\int\frac{d^{3}{\bf p}}{(2\pi)^{3}}
\left(\frac{s\mathbf{p}}{2|\mathbf{p}|^{2}}\right)\nonumber \\
 &  & \times\left[f_{\mathrm{FD}}(\gamma\beta|\mathbf{p}|-\gamma\beta\mathbf{v}\cdot\mathbf{p}-\beta\mu_{s})
 +f_{\mathrm{FD}}(\gamma\beta|\mathbf{p}|-\gamma\beta\mathbf{v}\cdot\mathbf{p}+\beta\mu_{s})\frac{}{}\right]\,,
\end{eqnarray}
where $f_{\mathrm{FD}}(y)\equiv1/(e^y+1)$ is the Fermi-Dirac distribution function.
Note that we have absorbed the zeroth-order contribution $\mathbf{j}_{s}^{(0)}(1)$ into $\mathbf{j}_{s}(1)$
and taken the limit $|\mathbf{v}|=0$ for $u^{\mu}=(\gamma,\gamma\mathbf{v})$ with
$\gamma=1/\sqrt{1-|\mathbf{v}|^{2}}$ to obtain $\mathbf{j}_{s}(2)$.
We see that $\mathbf{j}_{s}(1)$ comes from the momentum integration of
the fermion's velocity $\mathbf{p}/|\mathbf{p}|$ weighted by
the Fermi-Dirac distribution function in which the fermion's
energy is modified by the spin-vorticity coupling, while $\mathbf{j}_{s}(2)$
is from the magnetization due to the magnetic moment of the chiral
fermion which is given by $\hbar s\mathbf{p}/(2|\mathbf{p}|^{2})$
\cite{Chen:2014cla,Chen:2015gta,Kharzeev:2016sut}.

\subsection{Generation of spin polarization effect}
Since the space component of the axial current measures
the spin vector density in phase space \cite{Fang:2016vpj},
Eq. (\ref{j5-1-a}) tells that the spin is globally polarized along the vorticity direction.
However, this is a static result and does not tell us how the spin polarization
is generated from zero spin polarization.
To see the dynamical process of spin polarization generation,
we consider a system of massive fermions in a transient electromagnetic field.{\cite{Gao:2019znl}}
At the initial time $t=0$, there is no electromagnetic field, and the system is
unpolarized so that ${\pmb{\mathcal{A}}}=0$ and $\mathcal{F}\neq 0$.
The electromagnetic field comes up at the next moment and the evolution of ${\pmb{\mathcal{A}}}$
after an infinitesimal time interval is
\begin{eqnarray}
 \label{A-evolution}
\nabla_t {\pmb{\mathcal{A}}}
&=&- \frac{\hbar\, }{2m E_p }({\bf B}+{\bf E}\times {\bf v})
(  {\bf v}\cdot {\pmb \nabla} + E_p
\overleftarrow{\pmb \nabla}_x\cdot {\pmb \nabla}_p ) \mathcal{F}\,.
\end{eqnarray}
It is obvious that the polarization can be generated along the magnetic field due to an inhomogeneous distribution $\mathcal{F}$ in phase space. If there is no external electromagnetic field at $t=0$,
the kinetic equations for $\mathcal{F}$ and ${\pmb{\mathcal{A}}}$ will be decoupled from each other
so that the spin polarization became impossible from an unpolarized initial state.
However the self-consistent electromagnetic field always arises from the electric current
through Maxwell's equation $\partial_\mu F^{\mu\nu}= e^2 j^\nu$.
Rewriting Maxwell's equations as quadratic equations for the field tensor, i.e.~\cite{Vasak:1987um},
\begin{equation}
\partial_\lambda\partial^\lambda F_{\mu\nu}=e^2 \partial_\mu j_\nu - \partial_\nu j_\mu.
\end{equation}
We see clearly that it is the vorticity of $j_\mu$ instead of the current itself that induces the electromagnetic field tensor. The induced electromagnetic field make $\pmb{\mathcal{A}}$ and $\mathcal{F}$ be coupled together and generate the spin polarization along the vorticity direction. This provides a mechanism for generation of the spin polarization from vorticity in the current. { It deserves a future investigation this mechanism by numerical simulation.}

\section{Summary and Outlook}
In this review article, we give a brief overview on recent progress of the quantum kinetic theory based on Wigner functions. We focus on chiral and spin kinetic equations as well as various novel effects associated
with chirality and spin. We show that the relativistic quantum kinetic equation can be derived by disentangling the original quantum transport equation with the help of the semiclassical expansion in the reduced Planck constant $\hbar$.

For massless fermions, the system can be described by chiral Wigner functions which are composed of vector
and axial vector components of the Wigner function. According to the DWF theorem,
among four components of the chiral Wigner function, there is only one independent component satisfying one kinetic equation and one on-shell condition. Normally we choose the distribution function, the time-like component of the chiral Wigner function, as the independent component at each order in $\hbar$, while spatial components at an order of $\hbar$ can be expressed as functions of distribution functions at the same and lower orders.
For massive fermions, the primary 16 components of the Wigner function can be reduced into 4 independent components. One can choose the scalar and axial vector components as independent ones, which correspond to the particle distribution function and spin polarization vector. The four independent functions satisfy four coupled kinetic equations.
The anomalous transport coefficients such as those of CME, CVE, CSE and LPE can be derived naturally from Wigner functions. A mechanism for generation of spin polarization is proposed as a result of the kinetic equation for the spin polarization vector.

Since we restrict ourselves to the background field approximation and neglected quantum effects of the gauge field, we do not include particle collisions \cite{Chen:2015gta,Hidaka:2016yjf,Li:2019qkf,Carignano:2019zsh,Yang:2020hri,Weickgenannt:2020aaf,Wang:2020pej}.
A combination of the background field and quantum fluctuation in the Wigner function formalism
is a possible way to deal with particle collisions and transports in electromagnetic fields,
which is important to apply the Wigner function formalism to real systems in heavy-ion collisions,
a possible direction to go in the future.

\section*{Acknowledgments}

This work was supported in part by the National Natural Science Foundation of China under
Nos. 11890710, 11890713 and 11535012 and by the Strategic Priority Research Program
of Chinese Academy of Sciences under Grant No. XDB34030102.




\end{document}